\pdfoutput=1
\documentclass[12pt,a4paper]{article}
\usepackage{ifthen} 
\usepackage{booktabs}
\usepackage{multirow}
\usepackage{bm}
\newboolean{pdflatex}
\setboolean{pdflatex}{true} 

\newboolean{articletitles}
\setboolean{articletitles}{true} 

\newboolean{uprightparticles}
\setboolean{uprightparticles}{false} 

\newboolean{inbibliography}
\setboolean{inbibliography}{false} 

\usepackage{microtype}
\usepackage{lineno}  
\usepackage{xspace} 

\usepackage{graphicx}  
\usepackage{color}
\usepackage{colortbl}
\graphicspath{{./}} 

\usepackage{amsmath} 
\usepackage{amssymb}
\usepackage{amsfonts}
\usepackage{upgreek} 

\newcommand*\patchAmsMathEnvironmentForLineno[1]{%
\expandafter\let\csname old#1\expandafter\endcsname\csname #1\endcsname
\expandafter\let\csname oldend#1\expandafter\endcsname\csname
end#1\endcsname
 \renewenvironment{#1}%
   {\linenomath\csname old#1\endcsname}%
   {\csname oldend#1\endcsname\endlinenomath}%
}
\newcommand*\patchBothAmsMathEnvironmentsForLineno[1]{%
  \patchAmsMathEnvironmentForLineno{#1}%
  \patchAmsMathEnvironmentForLineno{#1*}%
}
\AtBeginDocument{%
\patchBothAmsMathEnvironmentsForLineno{equation}%
\patchBothAmsMathEnvironmentsForLineno{align}%
\patchBothAmsMathEnvironmentsForLineno{flalign}%
\patchBothAmsMathEnvironmentsForLineno{alignat}%
\patchBothAmsMathEnvironmentsForLineno{gather}%
\patchBothAmsMathEnvironmentsForLineno{multline}%
}

\usepackage{hyperref}    
\usepackage[all]{hypcap} 

\def\lhcb {\mbox{LHCb}\xspace}

\ifthenelse{\boolean{uprightparticles}}%
{

 \def\Pmu         {\ensuremath{\upmu}\xspace}

 \def\Pchi        {\ensuremath{\upchi}\xspace}                 
 \def\Ppsi        {\ensuremath{\uppsi}\xspace}

 \def\PDelta      {\ensuremath{\Delta}\xspace}                 
 \def\PXi      {\ensuremath{\Xi}\xspace}                 
 \def\PLambda      {\ensuremath{\Lambda}\xspace}                 
 \def\PSigma      {\ensuremath{\Sigma}\xspace}                 
 \def\POmega      {\ensuremath{\Omega}\xspace}                 
 \def\PUpsilon      {\ensuremath{\Upsilon}\xspace}                 
 
 \def\PB      {\ensuremath{\mathrm{B}}\xspace}                 
                  
 \def\PD      {\ensuremath{\mathrm{D}}\xspace}

 \def\PJ      {\ensuremath{\mathrm{J}}\xspace}                 
 \def\PK      {\ensuremath{\mathrm{K}}\xspace}

 \def\Pb      {\ensuremath{\mathrm{b}}\xspace}                 
 \def\Pc      {\ensuremath{\mathrm{c}}\xspace}

 \def\Pi      {\ensuremath{\mathrm{i}}\xspace}

}
{

 \def\Pmu         {\ensuremath{\mu}\xspace}

 \def\Pchi        {\ensuremath{\chi}\xspace}                 
 \def\Ppsi        {\ensuremath{\psi}\xspace}                 
                  
 \mathchardef\PDelta="7101
 \mathchardef\PXi="7104
 \mathchardef\PLambda="7103
 \mathchardef\PSigma="7106
 \mathchardef\POmega="710A
 \mathchardef\PUpsilon="7107
                  
 \def\PB      {\ensuremath{B}\xspace}                 
                  
 \def\PD      {\ensuremath{D}\xspace}

 \def\PJ      {\ensuremath{J}\xspace}                 
 \def\PK      {\ensuremath{K}\xspace}

 \def\Pb      {\ensuremath{b}\xspace}                 
 \def\Pc      {\ensuremath{c}\xspace}

 \def\Pi      {\ensuremath{i}\xspace}

}

\def\mup        {\ensuremath{\Pmu^+}\xspace}
\def\mun        {\ensuremath{\Pmu^-}\xspace} 

\def\cquark    {\ensuremath{\Pc}\xspace}
\def\cquarkbar {\ensuremath{\overline \cquark}\xspace}
\def\ccbar     {\ensuremath{\cquark\cquarkbar}\xspace}
\def\bquark    {\ensuremath{\Pb}\xspace}

  \def\Kbar  {\kern 0.2em\overline{\kern -0.2em \PK}{}\xspace}

  \def\Dbar    {\kern 0.2em\overline{\kern -0.2em \PD}{}\xspace}

\def\Bbar    {\ensuremath{\kern 0.18em\overline{\kern -0.18em \PB}{}}\xspace}

\def\jpsi     {\ensuremath{{\PJ\mskip -3mu/\mskip -2mu\Ppsi\mskip 2mu}}\xspace}
\def\psitwos  {\ensuremath{\Ppsi{(2S)}}\xspace}

  \def\Y#1S{\ensuremath{\PUpsilon{(#1S)}}\xspace}

\def\chic  {\ensuremath{\Pchi_{c}}\xspace}

\def\Lbar {\ensuremath{\kern 0.1em\overline{\kern -0.1em\PLambda}}\xspace}

\newcommand{\decay}[2]{\ensuremath{#1\!\to #2}\xspace}         

\def\to                 {\ensuremath{\rightarrow}\xspace}

\def\AT#1     {\ensuremath{A_{\mathrm{T}}^{#1}}\xspace}           

\def\C#1      {\ensuremath{\mathcal{C}_{#1}}\xspace}                       
\def\Cp#1     {\ensuremath{\mathcal{C}_{#1}^{'}}\xspace}                    
\def\Ceff#1   {\ensuremath{\mathcal{C}_{#1}^{\mathrm{(eff)}}}\xspace}        
\def\Cpeff#1  {\ensuremath{\mathcal{C}_{#1}^{'\mathrm{(eff)}}}\xspace}       
\def\Ope#1    {\ensuremath{\mathcal{O}_{#1}}\xspace}                       
\def\Opep#1   {\ensuremath{\mathcal{O}_{#1}^{'}}\xspace}                    

\newcommand{\tev}{\ifthenelse{\boolean{inbibliography}}{\ensuremath{~T\kern -0.05em eV}\xspace}{\ensuremath{\mathrm{\,Te\kern -0.1em V}}\xspace}}
\newcommand{\gev}{\ensuremath{\mathrm{\,Ge\kern -0.1em V}}\xspace}
\newcommand{\mev}{\ensuremath{\mathrm{\,Me\kern -0.1em V}}\xspace}
\newcommand{\kev}{\ensuremath{\mathrm{\,ke\kern -0.1em V}}\xspace}
\newcommand{\ev}{\ensuremath{\mathrm{\,e\kern -0.1em V}}\xspace}
\newcommand{\gevc}{\ensuremath{{\mathrm{\,Ge\kern -0.1em V\!/}c}}\xspace}
\newcommand{\mevc}{\ensuremath{{\mathrm{\,Me\kern -0.1em V\!/}c}}\xspace}
\newcommand{\gevcc}{\ensuremath{{\mathrm{\,Ge\kern -0.1em V\!/}c^2}}\xspace}
\newcommand{\gevgevcccc}{\ensuremath{{\mathrm{\,Ge\kern -0.1em V^2\!/}c^4}}\xspace}
\newcommand{\mevcc}{\ensuremath{{\mathrm{\,Me\kern -0.1em V\!/}c^2}}\xspace}

\def\mum  {\ensuremath{\,\upmu\rm m}\xspace}

\def\mub{\ensuremath{\rm \,\upmu b}\xspace}

\def\invfb   {\ensuremath{\mbox{\,fb}^{-1}}\xspace}

\def\gsim{{~\raise.15em\hbox{$>$}\kern-.85em
          \lower.35em\hbox{$\sim$}~}\xspace}
\def\lsim{{~\raise.15em\hbox{$<$}\kern-.85em
          \lower.35em\hbox{$\sim$}~}\xspace}

\def\pt         {\mbox{$p_{\rm T}$}\xspace}

\def\evtgen     {\mbox{\textsc{EvtGen}}\xspace}

\def\geant      {\mbox{\textsc{Geant4}}\xspace}

\def\photos     {\mbox{\textsc{Photos}}\xspace}

\def\pythia     {\mbox{\textsc{Pythia}}\xspace}

\def\tell1  {TELL1\xspace}
\def\ukl1   {UKL1\xspace}

\usepackage{cite} 
\usepackage{mciteplus}

\newcommand{\ourdecay}{$J/\psi \rightarrow \mu^+ \mu^-$}

\usepackage{lscape}

\begin{document}


\begin{titlepage}
\pagenumbering{roman}

\vspace*{-1.5cm}
\centerline{\large EUROPEAN ORGANIZATION FOR NUCLEAR RESEARCH (CERN)}
\vspace*{1.5cm}
\hspace*{-0.5cm}
\begin{tabular*}{\linewidth}{lc@{\extracolsep{\fill}}r}
\ifthenelse{\boolean{pdflatex}}
{\vspace*{-2.7cm}\mbox{\!\!\!\includegraphics[width=.14\textwidth]{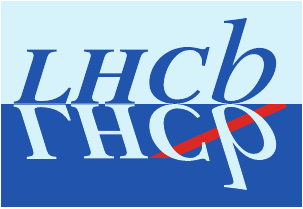}} & &}%
{\vspace*{-1.2cm}\mbox{\!\!\!\includegraphics[width=.12\textwidth]{lhcb-logo.eps}} & &}%
\\
 & & CERN-PH-EP-2013-130 \\  
 & & LHCb-PAPER-2013-008 \\  
 & & 24 July 2013\\ 
 & & \\
\end{tabular*}

\vspace*{4.0cm}

{\bf\boldmath\huge
\begin{center}
  Measurement of $J/\psi$ polarization \\
  in $pp$ collisions at $\sqrt{s}=7$ TeV
\end{center}
}

\vspace*{2.0cm}

\begin{center}
The LHCb collaboration\footnote{Authors are listed on the following pages.}
\end{center}

\vspace{\fill}

\begin{abstract}
  \noindent
\noindent An angular analysis of the decay \decay{\jpsi}{\mup\mun} is performed to measure
the polarization of prompt \jpsi mesons produced in $pp$ collisions at $\sqrt{s}=7 \tev$. The dataset corresponds to an integrated luminosity of 0.37~\invfb collected with the LHCb detector.
The measurement is presented as a function of transverse momentum, \pt,
and rapidity, $y$, of the \jpsi meson, in the kinematic region $2 < \pt <
15~\gevc $ and $2.0<y<4.5$.
  
\end{abstract}

\vspace*{2.0cm}

\begin{center}
  Published in Eur. Phys. J. C
\end{center}

\vspace{\fill}

{\footnotesize 
\centerline{\copyright~CERN on behalf of the \lhcb collaboration, license \href{http://creativecommons.org/licenses/by/3.0/}{CC-BY-3.0}.}}
\vspace*{2mm}

\end{titlepage}


\newpage
\setcounter{page}{2}
\mbox{~}
\newpage

\centerline{\large\bf LHCb collaboration}
\begin{flushleft}
\small
R.~Aaij$^{40}$, 
C.~Abellan~Beteta$^{35,n}$, 
B.~Adeva$^{36}$, 
M.~Adinolfi$^{45}$, 
C.~Adrover$^{6}$, 
A.~Affolder$^{51}$, 
Z.~Ajaltouni$^{5}$, 
J.~Albrecht$^{9}$, 
F.~Alessio$^{37}$, 
M.~Alexander$^{50}$
% ...

{\footnotesize \it
% ...
$ ^{5}$Clermont Universit\'{e}, Universit\'{e} Blaise Pascal, CNRS/IN2P3, LPC, Clermont-Ferrand, France\\
$ ^{6}$CPPM, Aix-Marseille Universit\'{e}, CNRS/IN2P3, Marseille, France\\
% ...
$ ^{9}$Fakult\"{a}t Physik, Technische Universit\"{a}t Dortmund, Dortmund, Germany\\
% ...
$ ^{35}$Universitat de Barcelona, Barcelona, Spain\\
$ ^{36}$Universidad de Santiago de Compostela, Santiago de Compostela, Spain\\
$ ^{37}$European Organization for Nuclear Research (CERN), Geneva, Switzerland\\
% ...
$ ^{40}$Nikhef National Institute for Subatomic Physics, Amsterdam, The Netherlands\\
% ...
$ ^{45}$H.H. Wills Physics Laboratory, University of Bristol, Bristol, United Kingdom\\
% ...
$ ^{50}$School of Physics and Astronomy, University of Glasgow, Glasgow, United Kingdom\\
$ ^{51}$Oliver Lodge Laboratory, University of Liverpool, Liverpool, United Kingdom\\
% ...
$ ^{n}$LIFAELS, La Salle, Universitat Ramon Llull, Barcelona, Spain\\
% ...
}
\end{flushleft}

\cleardoublepage

\renewcommand{\thefootnote}{\arabic{footnote}}
\setcounter{footnote}{0}

\pagestyle{plain} 
\setcounter{page}{1}
\pagenumbering{arabic}

\section{Introduction}
\label{intro}
Studies of \jpsi production in hadronic collisions provide powerful tests of
QCD.
In $pp$ collisions, quarkonium resonances can be produced directly, through feed-down from higher quarkonium states (such as the \psitwos or \chic resonances~\cite{Brambilla}), or via the
decay of \bquark~hadrons.
The first two production mechanisms are generically referred to as prompt production.
The mechanism for prompt production is not yet fully understood and none of
the available models adequately predicts the
observed dependence of the \jpsi production cross-section and
polarization on its transverse momentum \pt ~\cite{Brambilla}.
This paper describes the measurement of the polarization of the prompt \jpsi
component in $pp$ collisions at $\sqrt{s} = 7$ \tev, using the dimuon decay mode.
The measured polarization is subsequently used to update the LHCb measurement of the
cross-section given in Ref.~\cite{LHCb-PAPER-2011-003}. This improves the precision of the cross-section measurement significantly as the polarization and overall reconstruction efficiency are highly correlated.
\par
The three polarization states of a \jpsi vector meson are specified
in terms of a chosen coordinate system in the rest frame of the
meson.
This coordinate system is called the polarization frame and is
defined with respect to a particular polarization axis.
Defining the polarization axis to be the $Z$-axis, the $Y$-axis is chosen to be orthogonal
to the production plane (the plane containing the \jpsi momentum and the beam
axis) and the $X$-axis is oriented to create a right-handed coordinate system.
\par
Several polarization frame definitions can be found in the literature.
In the helicity frame~\cite{Jacob} the polarization axis coincides with the
flight direction of the \jpsi in the centre-of-mass frame of the colliding
hadrons.
In the Collins-Soper frame~\cite{CollinsSoper} the polarization axis is the direction of the relative velocity of the colliding beams in the \jpsi rest frame.
\par
The angular decay distribution, apart from a normalization factor, is
described by

\begin{equation}
  \frac{d^2N}{d\cos\theta\;d\phi}
  \propto 1+\lambda_\theta\cos^2\!\theta
  + \lambda_{\theta \phi}\sin2\theta\cos\phi
  + \lambda_\phi\sin^2\!\theta\cos2\phi,
  \label{theory1}
\end{equation}

\noindent
where $\theta$ is the polar angle between the direction of the positive lepton and the chosen polarization axis, and $\phi$ is the azimuthal angle, measured with
respect to the production plane. In this formalism, the polarization is
completely longitudinal if the set of polarization parameters
($\lambda_\theta$, $\lambda_{\theta \phi}$, $\lambda_{\phi}$) takes the values
$(-1, 0, 0)$ and it is completely transverse if it takes the values $(1, 0, 0)$.
In the zero polarization scenario the parameters are $(0, 0, 0)$.
In the general case, the values of
($\lambda_\theta$, $\lambda_{\theta \phi}$, $\lambda_{\phi}$) depend on the choice of the spin quantization frame and
different values can be consistent with the same underlying polarization
states.
However, the combination of parameters

\begin{equation}
  \lambda_{\mathrm{inv}} = \frac{\lambda_\theta+3\lambda_\phi}{1-\lambda_\phi}
  \label{invpar}
\end{equation}
is invariant under the choice of polarization frame~\cite{Faccioli1, Faccioli2}.
The natural polarization axis for the measurement is that where the lepton azimuthal angle distribution is symmetric
($\lambda_\phi = \lambda_{\theta \phi}=0$) and $\lambda_\theta$ is maximal~\cite{Faccioli3}.
\par
Several theoretical models are used to describe quarkonium production, predicting the values and the kinematic dependence of the cross-section and
polarization.
The colour-singlet model (CSM) at leading order \cite{Chang,Baier}
underestimates the \jpsi production cross-section by two orders of magnitude
\cite{LHCb-PAPER-2011-003,CDFresults_jpsi_cs} and predicts significant
transverse polarization.
Subsequent calculations at next-to-leading order and at
next-to-next-to-leading order change these predictions dramatically.
The cross-section prediction comes close to the observed values and the
polarization is expected to be large and longitudinal~\cite{Campbell, Artoisenet, Gong, Landsberg}.
Calculations performed in the framework of non-relativistic quantum
chromodynamics (NRQCD), where the \ccbar pair can be produced in colour-octet
states (color-octet model, COM \cite{Bodwin, Cho, Cho2}), can explain the
shape and magnitude of the measured cross-section as a function of \pt.
COM predicts a dependence of the \jpsi polarization on the \pt of the \jpsi meson.
In the low \pt region (\pt $<M(\jpsi)/c$ with $M(\jpsi)$ the mass of the \jpsi meson), where the gluon fusion process
dominates, a small longitudinal polarization is expected~\cite{Beneke1}.
For \mbox{\pt $\gg M(\jpsi)$}, where gluon fragmentation dominates,
the leading order predictions~\cite{Beneke2, Braaten} and next-to-leading order
calculations~\cite{Gong2} suggest a large transverse component of the
\jpsi polarization. 
\par
The polarization for inclusive \jpsi production (including the feed-down from higher charmonium states) in hadronic interactions has been
measured by several experiments at Fermilab~\cite{CDFresults_jpsi_pol},
Brookhaven~\cite{RHICresults} and DESY~\cite{HERAresults}.
The CDF experiment, in $p\overline{p}$ collisions at $\sqrt{s}=1.96$ TeV, measured a
small longitudinal \jpsi polarization, going to zero at small \pt.
This measurement is in disagreement with the COM calculations and does not support the conclusion that the colour-octet terms dominate the \jpsi
production in the high \pt region.
The PHENIX experiment measured the \jpsi polarization in $pp$ collisions at
$\sqrt{s}=200 \gev$, for \pt $<3$ \gevc.
The HERA-B experiment studied \jpsi polarization in 920 \gevc fixed target
proton-nucleus ($p$-$C$ and $p$-$W$) collisions.
The explored kinematic region is defined for \pt $<5.4$ \gevc and Feynman variable $x_{\mathrm{F}}$ between $-0.34$ and $0.14$.
Also in these cases a small longitudinal polarization is observed.
Recently, at the LHC, ALICE~\cite{ALICEresults} and CMS~\cite{Chatrchyan:2013cla} have measured the \jpsi polarization
in $pp$ collisions at $\sqrt{s} = 7$ \tev, in the kinematic ranges of $2<\pt<8\gevc$, $2.5<y<4.0$, and  $14<\pt<70~\gevc$, $\left|y\right|<1.2$, respectively. 
The ALICE collaboration finds a small longitudinal polarization vanishing at high values of \pt \footnote{In the ALICE measurement
the \jpsi from $b$ decays are also included.}, while the CMS results do not show evidence of large transverse or longitudinal polarizations.
\par
The analysis presented here is performed by fitting the efficiency-corrected
angular distribution of the data.
Given the forward geometry of the LHCb experiment, the polarization results are
presented in the helicity frame and, as a cross-check, in the Collins-Soper
frame.
The polarization is measured by performing a two-dimensional angular analysis
considering the distribution given in Eq.~(\ref{theory1}) and using an
unbinned maximum likelihood fit.
To evaluate the detector acceptance, reconstruction and trigger efficiency,
fully simulated events are used.
The measurement is performed in six bins of \jpsi transverse momentum and five
rapidity bins. The edges of the bins in \jpsi \pt and $y$ are defined respectively as \mbox{[2, 3, 4, 5, 7, 10, 15] \gevc} in \jpsi \pt and [2.0, 2.5, 3.0, 3.5, 4.0, 4.5] in \jpsi $y$.
\par
The remainder of the paper is organized as following. In Sec.~\ref{detsample} a brief description of the LHCb
detector and the data sample used for the analysis is given. In
Sec.~\ref{sec_sigsel} the signal selection is defined. In Sec.~\ref{polfit}
and Sec.~\ref{sec:Systematics} respectively, the fit procedure to the angular
distribution and the contributions to the systematic uncertainties on the
measurement are described. The results are presented in Sec.~\ref{results} and
in Sec.~\ref{crosssecref} the update of the \jpsi cross-section,
including the polarization effect, is described. Finally in
Sec.~\ref{conclusion} conclusions are drawn. 

\section{LHCb detector and data sample} \label{detsample}

The LHCb detector~\cite{Alves:2008zz} is a single-arm forward spectrometer
covering the pseudorapidity range $2 < \eta < 5$, designed for the study of
hadrons containing $b$ or $c$ quarks.
A right-handed Cartesian coordinate system is used, centred on the nominal $pp$ collision point with $z$ pointing downstream
along the nominal beam axis and $y$ pointing upwards.
The detector includes a high precision tracking system consisting of a
silicon-strip vertex detector surrounding the $pp$ interaction
region, a large-area silicon-strip detector located upstream of a dipole
magnet with a bending power of about 4~Tm, and three stations of silicon-strip
detectors and straw drift tubes placed downstream.
The combined tracking system provides momentum measurement with
relative uncertainty that varies from 0.4\% at 5\gevc to 0.6\% at 100\gevc,
and impact parameter resolution of 20\mum for
tracks with high \pt.
Charged hadrons are identified using two ring-imaging Cherenkov
detectors.  Photon, electron and hadron candidates are identified by a calorimeter system
consisting of scintillating-pad and pre-shower detectors, an electromagnetic
calorimeter and a hadronic calorimeter.  Muons are identified by a
system composed of alternating layers of iron and multiwire
proportional chambers~\cite{LHCb-DP-2012-002}.
\par
The trigger~\cite{LHCbTrigger} consists of a
hardware stage, based on information from the calorimeter and muon
systems, followed by a software stage, which applies a full event
reconstruction.
Candidate events are selected by the hardware trigger requiring the \pt of one muon to be larger than 1.48~\gevc, 
or the products of the \pt of the two muons to be larger than \mbox{1.68~$(\mathrm{Ge\kern -0.1em V\!/}c)^2$}.
In the subsequent software trigger~\cite{LHCbTrigger}, two tracks with $\pt>0.5$~$\gevc$ and momentum $p>6$~$\gevc$ are required to be identified as muons and the
invariant mass of the two muon tracks is required to be within $\pm 120 \mevcc$
of the nominal mass of the \jpsi meson \cite{PDG2012}.
The data used for this analysis correspond to an integrated luminosity of
0.37~\invfb of $pp$ collisions at a center-of-mass energy of $\sqrt{s}=
7$~TeV, collected by the LHCb experiment in the first half of 2011. The period
of data taking has been chosen to have uniform trigger conditions.
\par
In the simulation, $pp$ collisions are generated using
\pythia~6.4~\cite{Sjostrand:2006za} with a specific \lhcb
configuration~\cite{LHCb-PROC-2010-056}. Decays of hadronic particles
are described by \evtgen~\cite{Lange:2001uf}, in which final state
radiation is generated using \photos~\cite{Golonka:2005pn}. The
interaction of the generated particles with the detector and its
response are implemented using the \geant
toolkit~\cite{Allison:2006ve, *Agostinelli:2002hh} as described in
Ref.~\cite{LHCb-PROC-2011-006}.
The prompt charmonium production is simulated in \pythia according to the
leading order colour-singlet and colour-octet mechanisms.

\section{Signal selection}
\label{sec_sigsel}
The selection requires that at least one primary vertex is reconstructed in
the event.
Candidate \jpsi mesons are formed from pairs of opposite-sign tracks
reconstructed in the tracking system.
Each track is required to have $p_\mathrm{T}>0.75$ \gevc and to be identified as a
muon.
The two muons must originate from a common vertex and the $\chi^2$
probability of the vertex fit must be greater than 0.5\%.
\begin{figure}[!t]
  \begin{minipage}[b]{7.5cm}
    \centering
    \includegraphics[width=7.5cm]{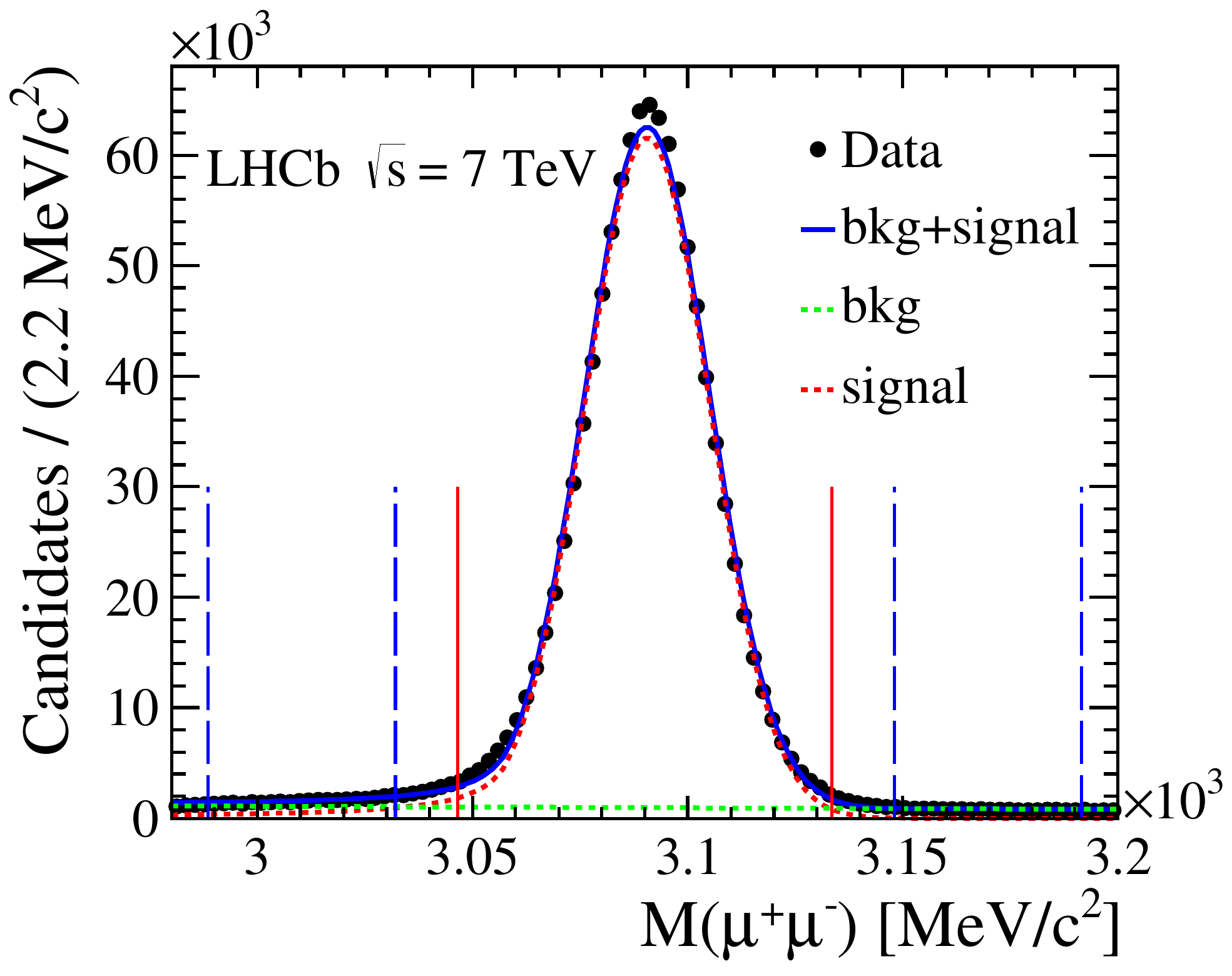}
  \end{minipage}
  \hspace{5mm}
  \begin{minipage}[b]{7.5cm}
    \centering
    \includegraphics[width=7.5cm]{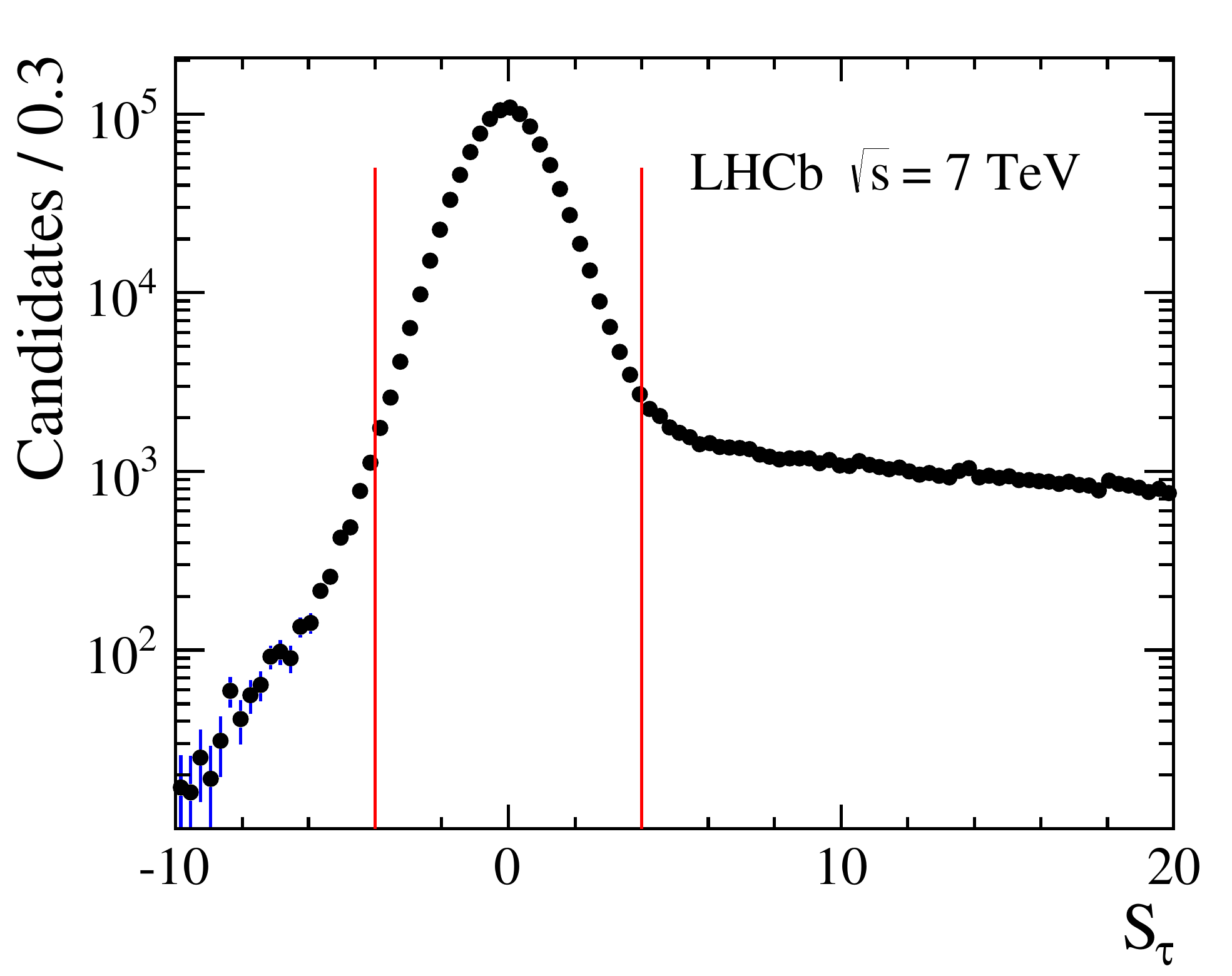}
  \end{minipage}
  \caption{
\small{(${\it Left}$)
 Invariant mass distribution of muon pairs
    passing the selection criteria.
    In the plot, \jpsi candidates are required to have 5
    $<$ \pt $<$ 7 \gevc and $3.0<y<3.5$. The solid (dashed) vertical
lines indicate the signal (sideband) regions.
    (${\it Right}$) Pseudo decay-time significance ($S_{\tau}$) distribution for background subtracted 
   \jpsi candidates in the same kinematic bin. The solid vertical lines indicate the $S_\tau$ selection region.
    The right tail of the distribution is due to \jpsi production through the decay of $b$ hadrons.
   }\label{fig:mass}}
\end{figure}

In Fig.~\ref{fig:mass} (left), the invariant mass
distribution of \jpsi candidates for \mbox{5 $<$ \pt $<$ 7 \gevc} and
\mbox{$3.0<y<3.5$} is shown as an example.
A fit to the mass distribution has been performed using a Crystal Ball function
\cite{Skwarnicki:1986xj} for the signal and a linear function
for the background, whose origin is combinatorial.
The Crystal Ball parameter describing the threshold of the radiative tail is fixed to the value obtained in the simulation. The Crystal Ball peak position and resolution determined in the fit shown in Fig.~\ref{fig:mass} (left) are respectively \mbox{$\mu =  3090.5$ \mevcc} and \mbox{$\sigma  =  14.6$ \mevcc}.
The signal region is defined as $\left[\mu-3\sigma,\;\mu+3\sigma\right]$ and the two sideband regions as
$\left[\mu-7\sigma,\mu-4\sigma\right]$ and $\left[\mu+4\sigma,\mu+7\sigma\right]$ in the mass distribution.
\par
Prompt $J/\psi$ mesons and \jpsi mesons from $b$-hadron decays can be discriminated by the pseudo-decay-time $\tau$, which is defined as: 
\begin{equation}\label{tzsdefinition}
  \tau = \frac{(z_{\jpsi} - z_{\mathrm{\,PV}})M(\jpsi)}{p_{z}}\;,
\end{equation}
\noindent
where $z_{\jpsi}$ and $z_{\mathrm{PV}}$ are the positions of
the \jpsi decay vertex and the associated primary vertex along the $z$-axis, $M(\jpsi)$ is
the nominal \jpsi mass and $p_{z}$ is the measured $z$ component of the \jpsi momentum in the center-of-mass frame of the $pp$ collision.
For events with several primary vertices, the one which is closest to the \jpsi vertex is used.
The uncertainty $\sigma_\tau$ is calculated for each candidate using the measured covariance matrix of
$z_{\jpsi}$ and $p_{z}$ and the uncertainty of $z_{\mathrm{PV}}$. The bias induced by not refitting the primary vertex removing the two tracks from the reconstructed \jpsi meson is found to be negligible \cite{LHCb-PAPER-2011-003}. The pseudo decay-time significance $S_\tau$ is defined as $S_{\tau} = \tau/\sigma_\tau$. In order to suppress the component of \jpsi mesons from $b$-hadron decays, it is required that $\vert S_{\tau} \vert < 4$. With this requirement, the fraction of \jpsi from $b$-hadron decays reduces from about 15\% to about 3\%.
The distribution of the pseudo-decay-time significance in one kinematic bin is shown in Fig.~\ref{fig:mass} (right).

\section{Polarization fit}
\label{polfit}
\begin{figure}[!t]
  \begin{minipage}[b]{7.5cm}
    \centering
    \includegraphics[width=7.5cm]{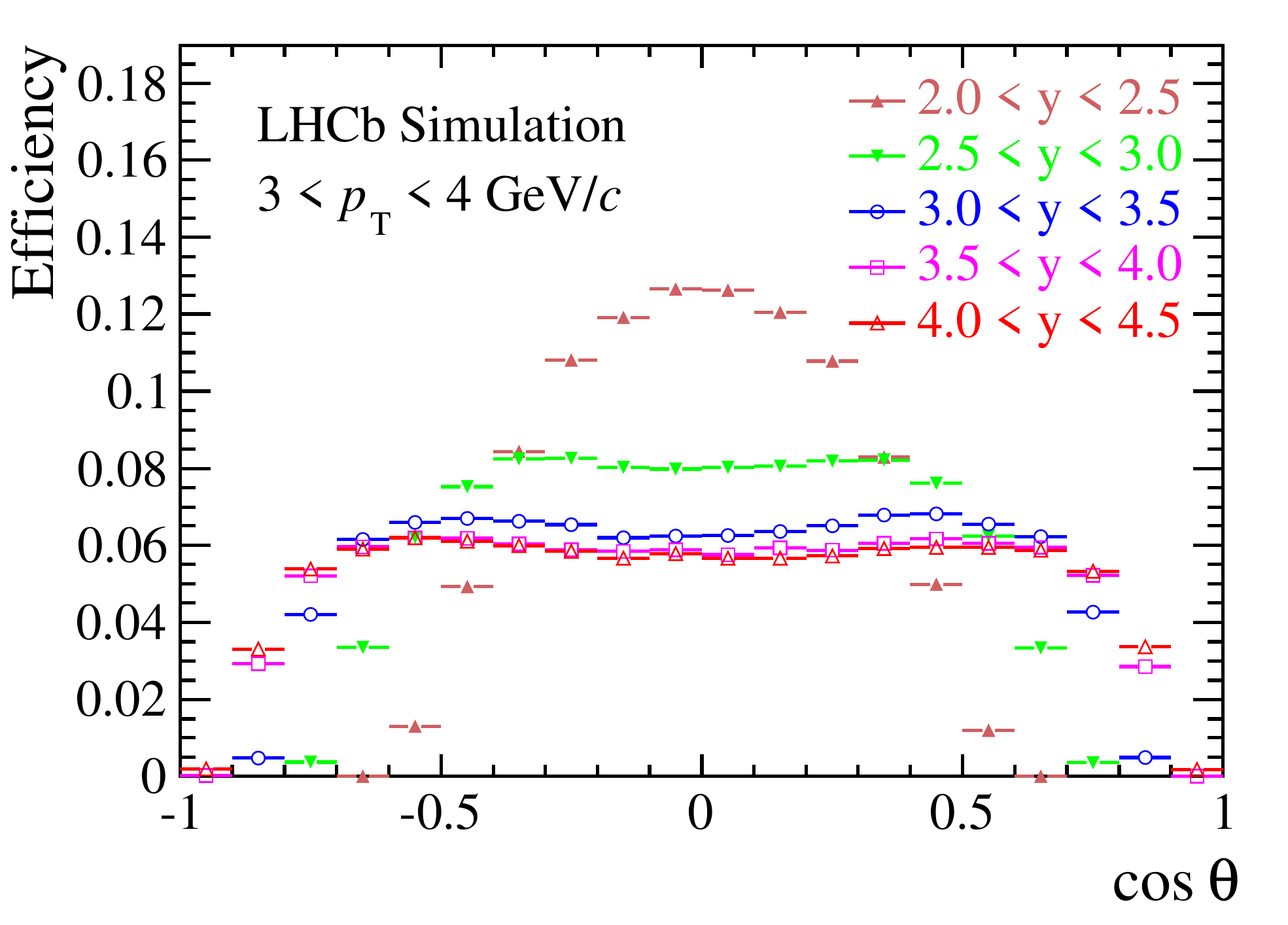}
  \end{minipage}
  \hfill
  \begin{minipage}[b]{7.5cm}
    \centering
    \includegraphics[width=7.5cm]{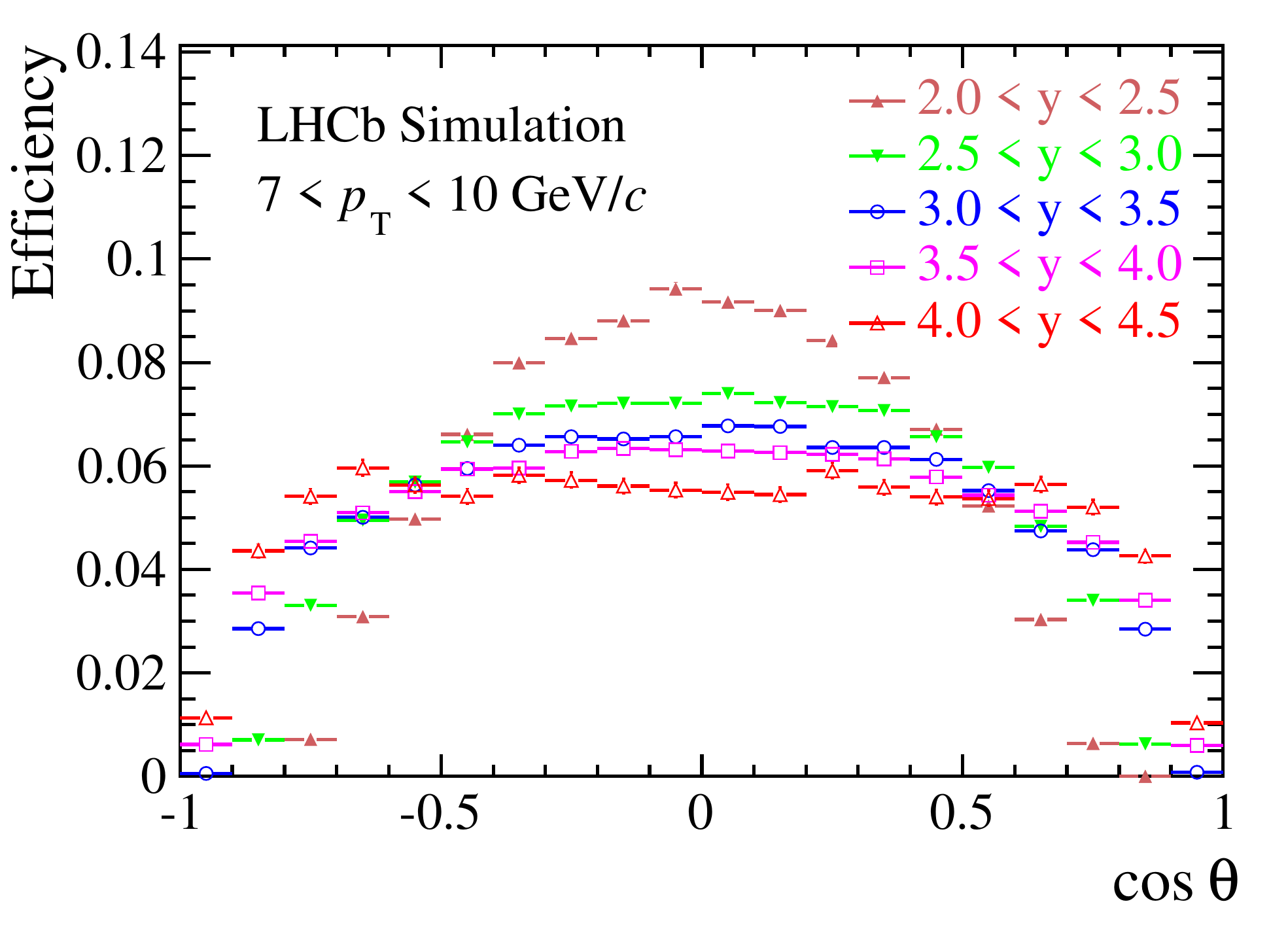}
  \end{minipage}
  \vfill
  \begin{minipage}[b]{7.5cm}
    \centering
    \includegraphics[width=7.5cm]{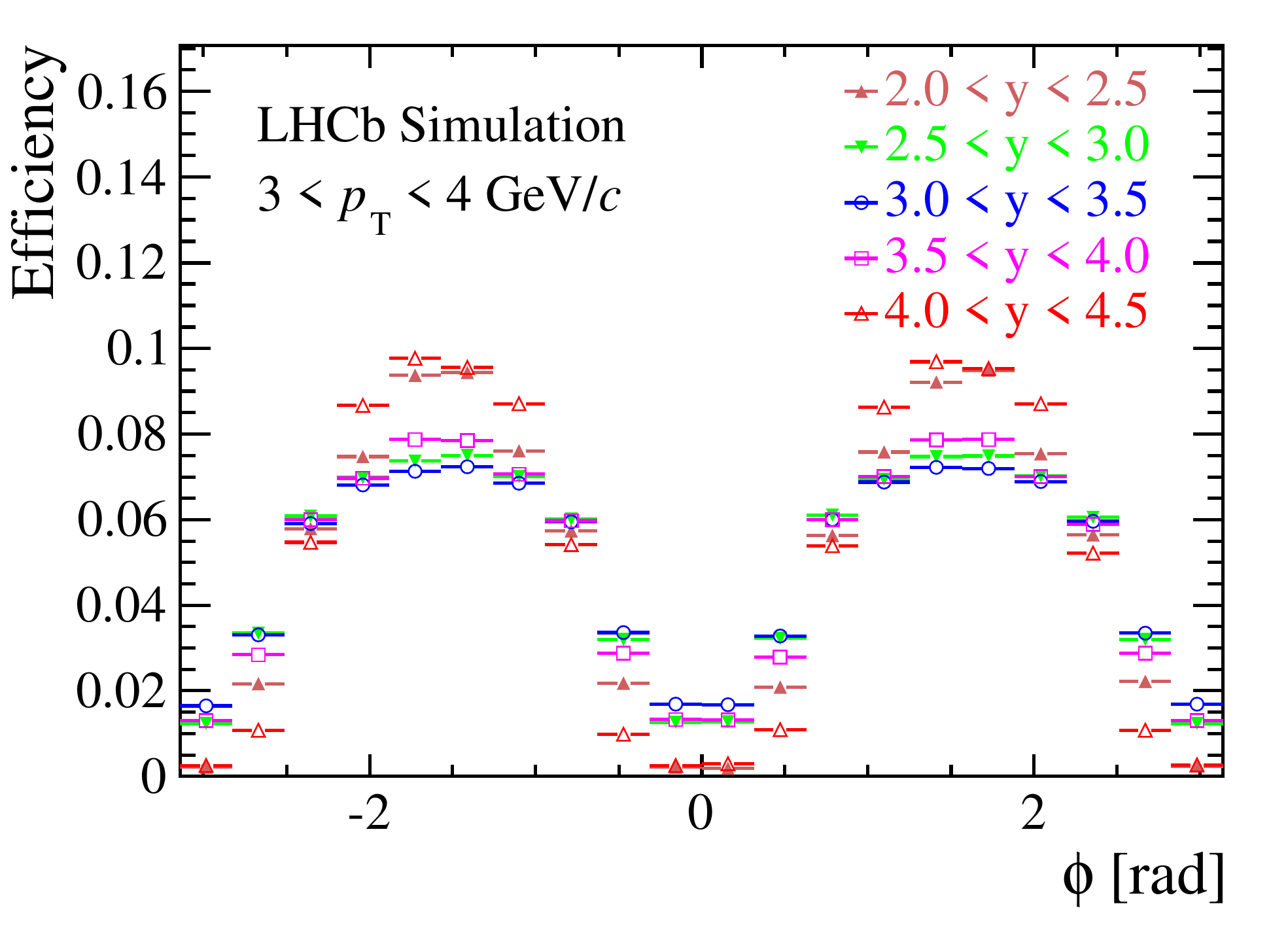}
  \end{minipage}
  \hfill
  \begin{minipage}[b]{7.5cm}
    \centering
    \includegraphics[width=7.5cm]{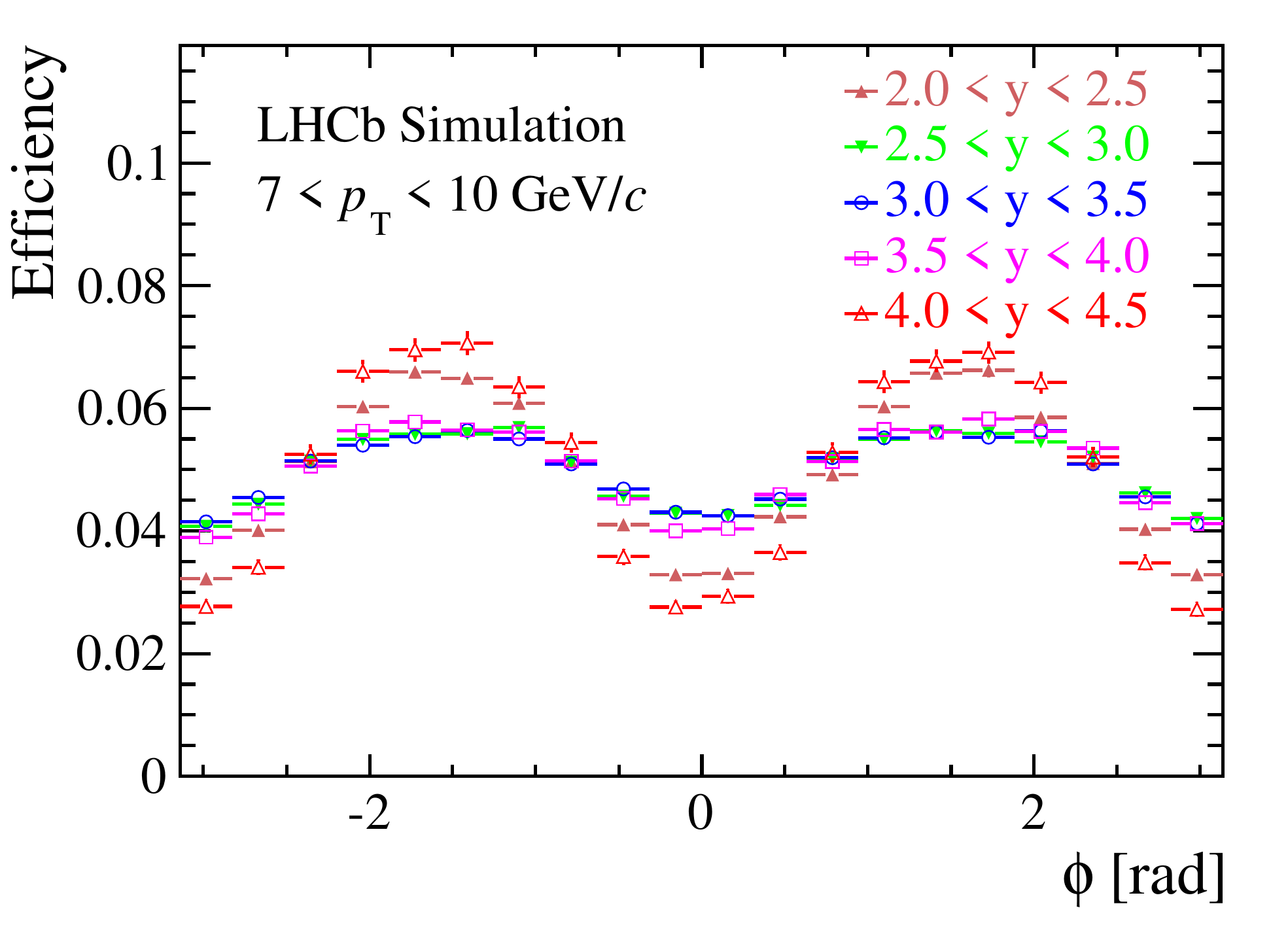}
  \end{minipage}
  \caption{\small{Global efficiency (area normalized to unity) as a
      function of (\textit{top}) $\cos \theta$ and (\textit{bottom}) $\phi$
      for (\textit{left}) 3 $<$ \pt $<$ 4 \gevc and for (\textit{right}) 7 $<$ \pt $<$ 10 \gevc of \jpsi mesons in the helicity frame.
      The efficiency is determined from simulation.}
    \label{fig:acceptance}}
\end{figure}

The polarization parameters are determined from a fit to the angular
distribution ($\cos\theta,\phi$) of the \ourdecay~decay.
The knowledge of the efficiency as a function of the angular
variables ($\cos\theta,\phi$) is crucial for the analysis.
The detection efficiency $\epsilon$ includes geometrical,
detection and trigger efficiencies and is obtained from
a sample of simulated unpolarized \jpsi mesons decaying in the \decay{\jpsi}{\mup\mun} channel, 
where the events are divided in bins of \pt and $y$ of the \jpsi meson.
The efficiency is studied as a function of four kinematic variables:
\pt and $y$ of the \jpsi meson, and $\cos \theta$ and $\phi$ of the positive
muon.
As an example, Fig.~\ref{fig:acceptance} shows the efficiency as a function of
$\cos\theta$ (integrated over $\phi$) and $\phi$ (integrated over
$\cos\theta$) respectively, for two different bins of \pt and all five bins of
$y$.
The efficiency is lower for $\cos\theta \approx \pm1$, as one of the two muons
in this case has a small momentum in the center-of-mass frame of the $pp$
collision and is often bent out of the detector acceptance by the dipole field
of the magnet.
The efficiency is also lower for $\vert\phi\vert \approx 0$ or $\pi$,
because one of the two muons often escapes the LHCb detector acceptance.

To fit the angular distribution in Eq.~(\ref{theory1}), a maximum
likelihood (ML) approach is used.
The logarithm of the likelihood function, for data in each \pt and $y$
bin, is defined as

\begin{eqnarray}\label{likelihood}
    \log L&=&\sum^{N_{\mathrm{tot}}}_{i=1}w_i\times
    \log\left[\frac{P(\cos\theta_{i},\phi_{i}\vert\lambda_{\theta},
        \lambda_{\theta\phi}, \lambda_{\phi})\;\epsilon(\cos\theta_{i},\phi_{i})}{N(\lambda_{\theta},
        \lambda_{\theta\phi}, \lambda_{\phi})}\right]\\
        &=&\sum^{N_{\mathrm{tot}}}_{i=1}w_i\times
    \log\left[\frac{P(\cos\theta_{i},\phi_{i}\vert\lambda_{\theta},
        \lambda_{\theta\phi}, \lambda_{\phi})}{N(\lambda_{\theta},
        \lambda_{\theta\phi}, \lambda_{\phi})}\right] + \sum^{N_{\mathrm{tot}}}_{i=1}w_{i}\times
    \log\left[\epsilon(\cos\theta_{i},\phi_{i})\right]\;, \label{likelihood2}
  \end{eqnarray}
\noindent

\noindent where $P(\cos\theta_{i},\phi_{i}\vert\lambda_{\theta},
        \lambda_{\theta\phi}, \lambda_{\phi}) = 1+\lambda_\theta \cos^2 \theta_{i}
  + \lambda_{\theta \phi}\sin 2\theta_{i} \cos \phi_{i}
  + \lambda_\phi\sin^2 \theta_{i} \cos 2\phi_{i}$, $w_i$ are weighting
factors and the index $i$ runs over the number of the candidates, $N_\mathrm{tot}$. The second sum in Eq.~(\ref{likelihood2}) can be ignored in the fit as it has no dependence on the polarization parameters.
$N(\lambda_{\theta}, \lambda_{\theta\phi}, \lambda_{\phi})$ is a normalization
integral, defined as
\begin{equation}\label{Normalization1}
  N(\lambda_{\theta},\lambda_{\theta\phi},\lambda_{\phi}) =
  \int d\Omega P(\cos\theta,\phi\vert\lambda_{\theta},
        \lambda_{\theta\phi}, \lambda_{\phi})\times\epsilon(\cos\theta,\phi) \;.
\end{equation}
\noindent
In the simulation where \jpsi mesons are generated unpolarized, the $(\cos\theta$, $\phi)$ two-dimensional distribution of selected candidates is the same as the efficiency $\epsilon(\cos\theta,\phi)$, so Eq. (\ref{Normalization1}) can be evaluated by summing $P(\cos\theta_{i},\phi_{i}\vert\lambda_{\theta}, \lambda_{\theta\phi}, \lambda_{\phi})$ over the \jpsi candidates in the simulated sample. The normalization $N(\lambda_{\theta}, \lambda_{\theta\phi}, \lambda_{\phi})$ depends on all three polarization parameters.
The weighting factor $w_i$ is chosen to be \mbox{$+1$ ($-1$)} if a candidate falls in
the signal region (sideband regions) shown in Fig.~\ref{fig:mass}.
In this way the background component in the signal window is subtracted on a
statistical basis.\footnote{The signal window and the sum of the sideband
  regions have the same width.}
For this procedure it is assumed that the angular distribution
$(\cos\theta,\phi)$ of background events in the signal region is similar to
that of the events in sideband regions, and that the mass distribution of the background is
approximately linear.
\par
The method used for the measurement of the polarization is tested by
measuring the \jpsi polarization in two simulated samples with a fully
transverse and fully longitudinal polarization, respectively.
In both cases the results reproduce the simulation input within the
statistical sensitivity.
\par
To evaluate the normalization function $N(\lambda_{\theta},
\lambda_{\theta\phi}, \lambda_{\phi})$ on the simulated sample of unpolarized \jpsi mesons, we rely on the correct simulation of the efficiency.
In order to cross check the reliability of the efficiency obtained from the simulation, the control-channel
\mbox{$B^+ \rightarrow \jpsi \, K^+$} is studied.
The choice of this channel is motivated by the fact that, due to angular
momentum conservation, the \jpsi must be longitudinally polarized and any
difference between the angular distributions measured in data and in the
simulation must be due to inaccuracies in the simulation.
\par
\begin{figure}[!t]
  \begin{minipage}[t]{0.5\textwidth}
    \centering
    \includegraphics[width=1.0\textwidth]{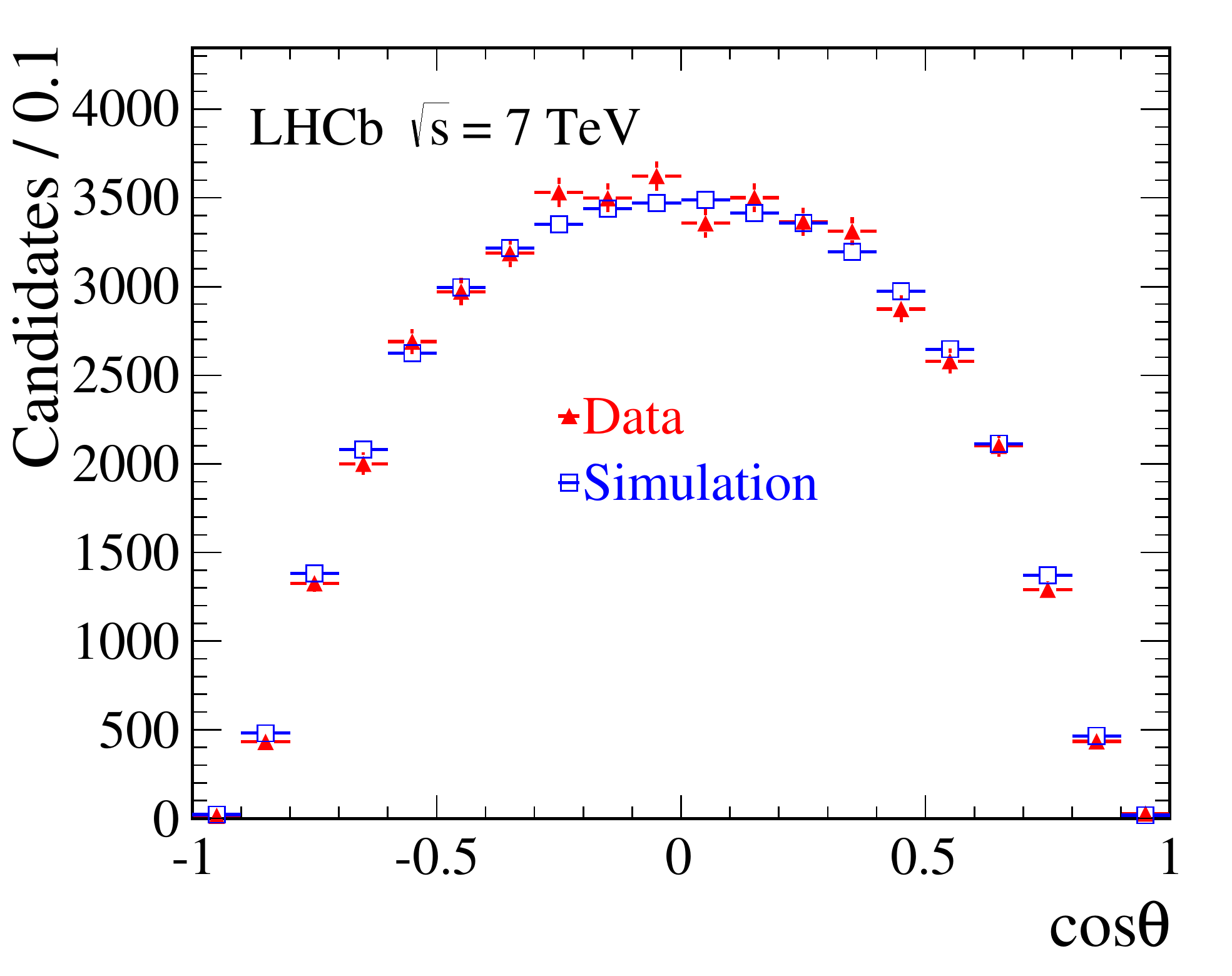}
  \end{minipage}
  \begin{minipage}[t]{0.5\textwidth}
    \centering
    \includegraphics[width=1.0\textwidth]{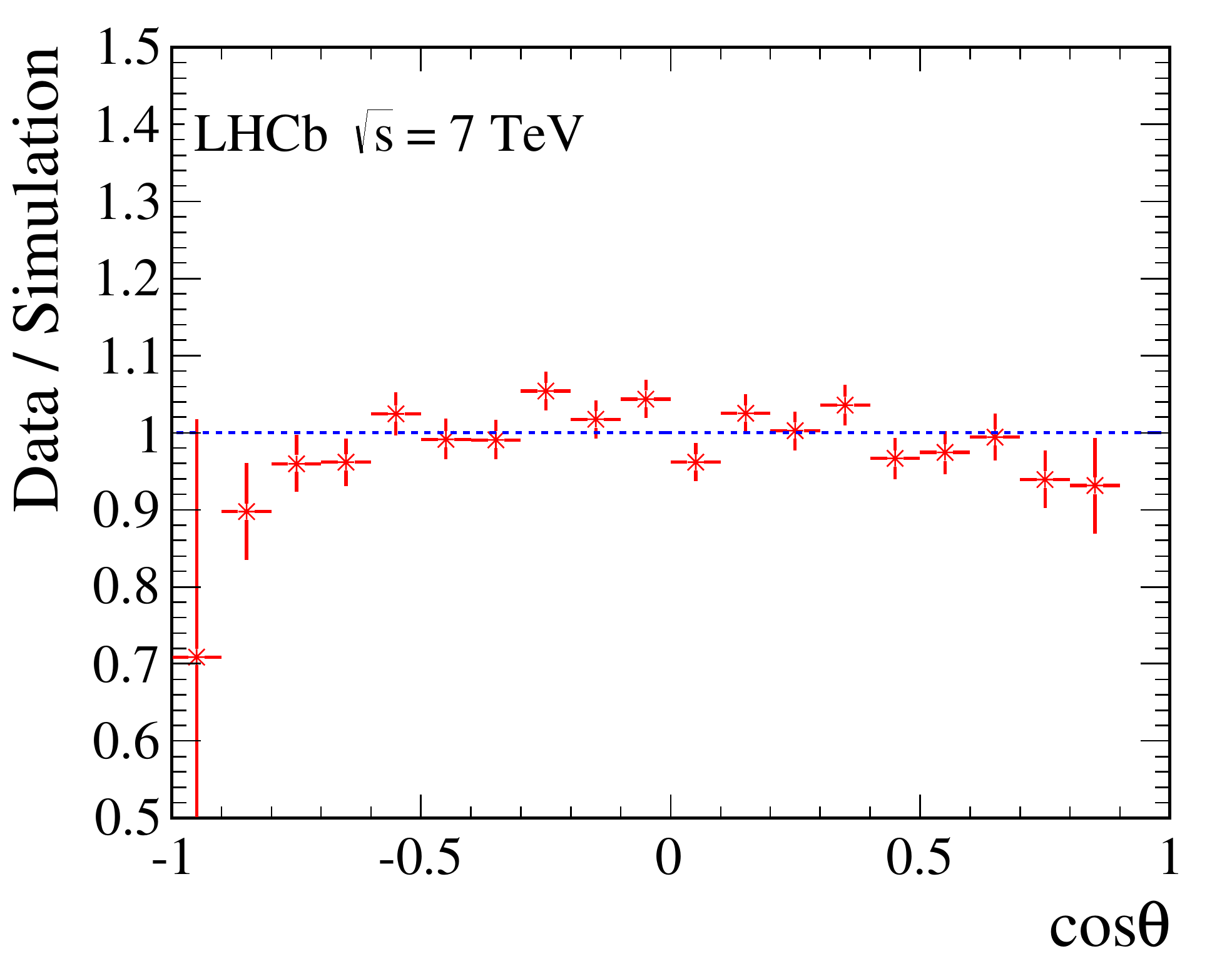}
  \end{minipage}
  \caption{\small{(\textit{Left}) Distributions of $\cos\theta$ in the helicity frame for \jpsi mesons from
    $B^+ \rightarrow \jpsi K^+$ decays in data (open circles) and simulated
    sample (open squares) after the weighting based on the $B^{+}$ and \jpsi kinematics and (\textit{right}) their ratio.}
  \label{fig:B2JpsiKMAngleB}}
\end{figure}

To compare the kinematic variables of the muons in data and simulation, a
first weighting procedure is applied to the simulated sample to
reproduce the $B^{+}$ and \jpsi kinematics in the data.
In Fig.~\ref{fig:B2JpsiKMAngleB}, $\cos \theta$ distributions for \mbox{$B^+ \rightarrow \jpsi \, K^+$} candidates
for data and simulation are shown, as well as their ratio.
A small difference between the distributions for data and simulation is observed, which
is attributed to an overestimation of the efficiency in the simulation for
candidates with values of $\left|\cos\theta\right|\approx1$.
To correct for the acceptance difference, an additional event weighting is applied
where the weighting factors are obtained by comparing the two-dimensional muon
\pt and $y$ distribution in the center-of-mass frame of $pp$ collisions in data and simulation.
This weighting corrects for the observed disagreement in the $\cos \theta$
distribution.
The \mbox{weights} as a function of muon \pt and $y$ obtained from the
\mbox{$B^+ \rightarrow \jpsi \, K^+$} sample are subsequently applied in the same way to
the simulated prompt \jpsi sample, which is used to determine the efficiency for
the polarization measurement.

\section{Systematic uncertainties}
\label{sec:Systematics}

The largest systematic uncertainty is related to the determination of the 
efficiency and to the weighting procedure used to correct the
simulation, using the $B^{+}\rightarrow J/\psi \, K^{+}$ control channel.
The weighting procedure is performed in bins of \pt and $y$ of the two muons 
and, due to the limited number of candidates in the control channel, the
statistical uncertainties of the correction factors are sizeable (from 1.3\%
up to 25\%, depending on the bin).
To propagate these uncertainties to the polarization results, the following
procedure is used.
For each muon ($\pt, y$) bin, the weight is changed by one standard deviation, 
leaving all other weights at their nominal values.
This new set of weights is used to redetermine the detector efficiency and
then perform a new fit of the polarization parameters.
The difference of the obtained parameters with respect to the nominal
polarization result is considered as the contribution of this muon $(\pt,y)$ bin to
the uncertainty.
The total systematic uncertainty is obtained by summing all these independent
contributions in quadrature.
In the helicity frame, the average absolute uncertainty over all the \jpsi $(\pt,y)$ bins due to this effect is
0.067 on $\lambda_\theta$.
\par
Concerning the background subtraction, the choice of the sidebands and
the background model are checked.
A systematic uncertainty is evaluated by comparing the nominal results for
the polarization parameters, and those obtained using only the left or the
right sideband, or changing the background fit function (as alternatives to
the linear function, exponential and polynomial functions are used).
In both cases the maximum variation with respect to the nominal result is
assigned as systematic uncertainty.
Typically, the absolute size of this effect is 0.012 on $\lambda_\theta$ for 
$\pt > 5 \gevc$.
\par
The effect of the ($\pt, y$) binning for the \jpsi meson could also introduce an
uncertainty, due to the difference of the \jpsi kinematic distributions
between data and simulation within the bins.
To investigate this effect, each bin is divided in four sub-bins
($2\times 2$) and the polarization parameters are calculated in each sub-bin.
The weighted average of the results in the four sub-bins is compared with the
nominal result and the difference is quoted as the systematic uncertainty.
As expected, this effect is particularly important in the rapidity range near
the LHCb acceptance boundaries, where the efficiency has a strong dependence on
the kinematic properties of the \jpsi meson.
It however depends on \pt only weakly and the average effect on
$\lambda_\theta$ is 0.018 (absolute).
\par
Two systematic uncertainties related to the cut on the \jpsi decay time
significance are evaluated.
The first is due to the residual \jpsi candidates from $b$-hadron decays,
3\% on average and up to 5\% in the highest \pt bins, that potentially have different polarization.
The second is due to the efficiency difference in the $S_{\tau}$
requirement in data and simulation.
The average size of these effects, over the \jpsi $(\pt, y)$, is 0.012.
\par
The limited number of events in the simulation sample, used to evaluate the
normalization integrals of Eq.~(\ref{Normalization1}), is a source of
uncertainty.
This effect is evaluated by simulating a large number of pseudo-experiments
and the average absolute size is 0.015.
\par
Finally, the procedure used to statistically subtract the background introduces
a statistical uncertainty, not included in the standard likelihood
maximization uncertainty.
A detailed investigation shows that it represents a tiny correction to the
nominal statistical uncertainty, reported in
Tables~\ref{tab:PolarizationResultHX} and~\ref{tab:PolarizationResultCS}.
\par
The main contributions to the systematic uncertainties on $\lambda _{\theta}$
are summarized in Table~\ref{tab:systematic} for the helicity and the
Collins-Soper frames.
While all uncertainties are evaluated for every \pt and $y$ bin separately, we
quote for the individual contributions only the average, minimum and maximum
values.
The systematic uncertainties on $\lambda_{\theta\phi}$ and $\lambda_\phi$ are
similar to each other and a factor two lower than those for $\lambda_\theta$.
Apart from the binning and the simulation sample size effects, the
uncertainties of adjacent kinematic bins are strongly correlated.
\par
To quote the global systematic uncertainty
(Tables~\ref{tab:PolarizationResultHX} and~\ref{tab:PolarizationResultCS}) in each kinematic bin of the \jpsi meson,
the different contributions for each bin are considered to be uncorrelated and are added in
quadrature.

\begin{table}
  \caption{\small
    Main contributions to the absolute systematic uncertainty on the
    parameter $\lambda_\theta$ in the helicity and Collins-Soper frames.
    While the systematic uncertainties are evaluated separately for all \pt
    and $y$ bins, we give here only the average, the minimum and the maximum
    values of all bins.}
  \label{tab:systematic}
  \begin{center}
    \begin{tabular}{lcc}
        \toprule[1.pt]
      \multirow{2}{*}{Source} & helicity frame &  Collins-Soper frame \\
      & average (min. -- max.) &  average (min. -- max.) \\
      \midrule[1.pt]
      Acceptance     & 0.067 (0.045 -- 0.173) & 0.044 (0.025 -- 0.185) \\
      Binning effect & 0.018 (0.001 -- 0.165) & 0.016 (0.001 -- 0.129) \\
      Simulation sample size & 0.015 (0.005 -- 0.127) & 0.015 (0.007 -- 0.170) \\
      Sideband subtraction & 0.016 (0.001 -- 0.099) & 0.029 (0.001 -- 0.183) \\
      \bquark-hadron contamination & 0.012 (0.002 -- 0.019) & 0.006 (0.002 -- 0.029) \\
      \bottomrule[1.pt]
    \end{tabular}
  \end{center}
\end{table}

\newcommand{\hd}{\hphantom{0}}
\section{Results}
\label{results}
\begin{figure}[t]
  \begin{minipage}[t]{0.5\textwidth}
    \centering
    \includegraphics[width=1.0\textwidth]{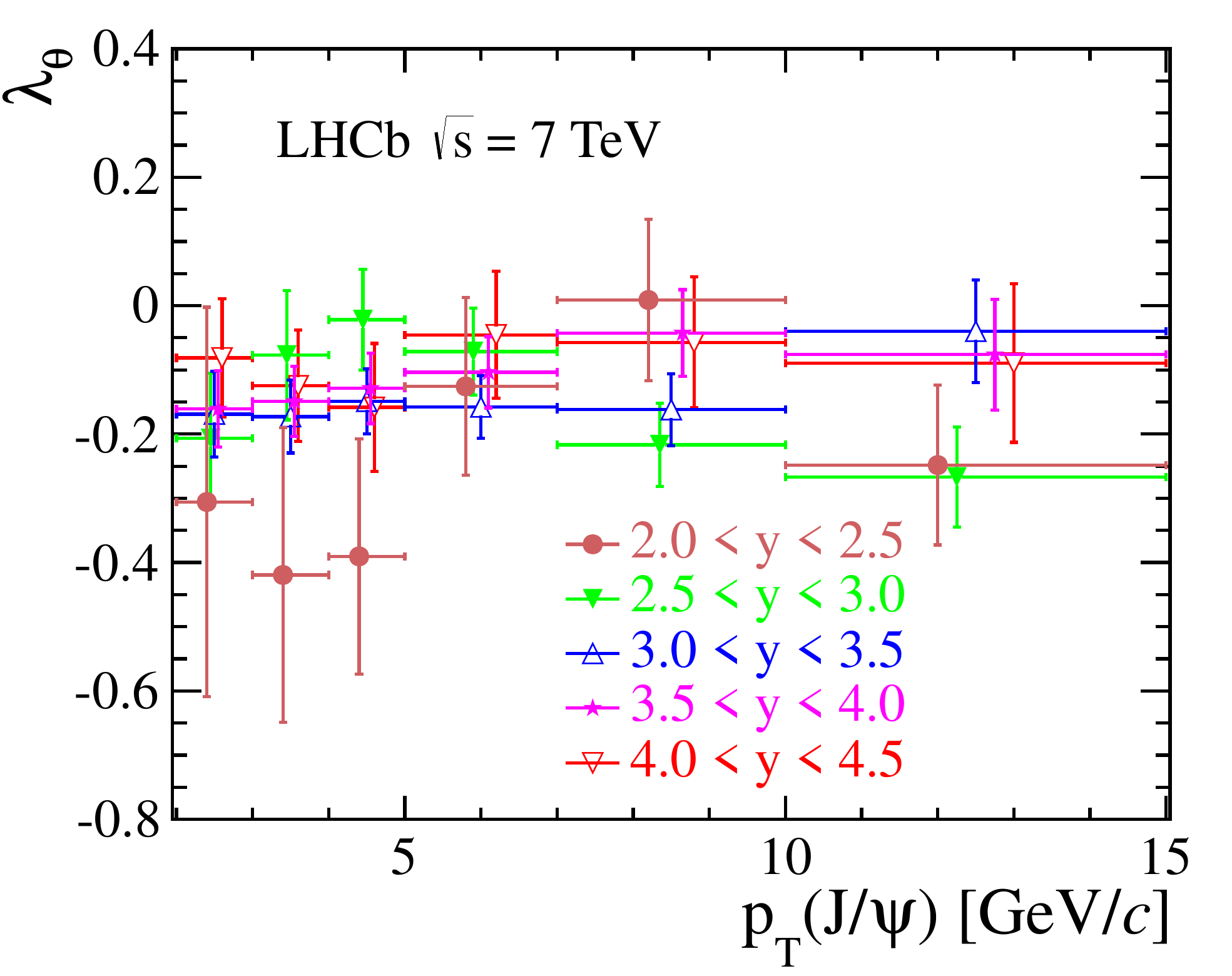}
  \end{minipage}
  \begin{minipage}[t]{0.5\textwidth}
    \centering
    \includegraphics[width=1.0\textwidth]{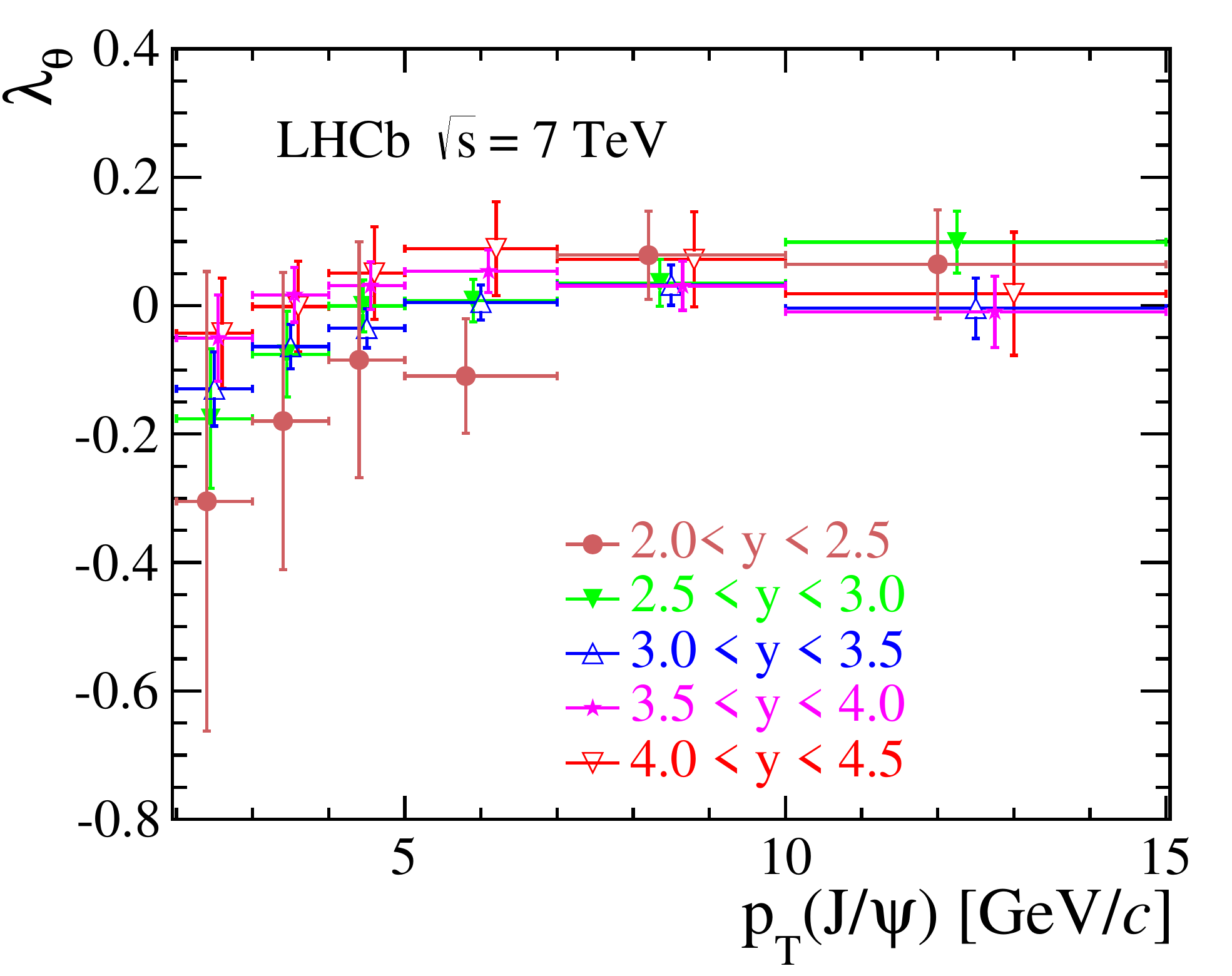}
  \end{minipage}
  \caption{\small
    Measurements of $\lambda_{\theta}$ in bins of \pt for five rapidity
    bins in (\textit{left}) the helicity frame and (\textit{right}) the
    Collins-Soper frame.
    The error bars represent the statistical and systematic uncertainties
    added in quadrature.
    The data points are shifted slightly horizontally for different rapidities
    to improve visibility.}
  \label{LambdaTheta}
\end{figure}

The fit results for the three parameters $\lambda_{\theta}$,
$\lambda_{\theta\phi}$ and $\lambda_{\phi}$, with their uncertainties, are
reported in Tables~\ref{tab:PolarizationResultHX} and~\ref{tab:PolarizationResultCS} for the helicity frame and the
Collins-Soper frame, respectively.
The parameter $\lambda_{\theta}$ is also shown in Fig.~\ref{LambdaTheta} as a
function of the \pt of the \jpsi meson, for different $y$ bins.

\par
The polarization parameters $\lambda_\phi$ and $\lambda_{\theta \phi}$ in the
helicity frame are consistent with zero within the uncertainties.
Following the discussion in Sec.\ref{intro}, the helicity frame represents
the natural frame for the polarization measurement in our experiment and the
measured $\lambda_\theta$ parameter is an indicator of the \jpsi polarization,
since it is equal to the invariant parameter defined in Eq.~(\ref{invpar}).
\par
The measured value of $\lambda_\theta$ shows a small longitudinal
polarization.
A weighted average is calculated over all the $(\pt , y)$ bins, where the
weights are chosen according to the number of events in each bin in the data
sample.
The average is $\lambda_\theta= -0.145 \pm 0.027$.
The uncertainty is statistical and systematic uncertainties added in quadrature.
Since the correlations of the systematic uncertainties are observed to be relevant 
only between adjacent kinematic bins, when quoting the average uncertainty, we 
assume the different kinematic bins are uncorrelated, apart from the
adjacent ones, which we treat fully correlated.
\par
A cross-check of the results is performed by repeating the measurement in
the Collins-Soper reference frame (see Sec.~\ref{intro}).
As LHCb is a forward detector, the Collins-Soper and helicity frames are
kinematically quite similar, especially in the low \pt and $y$ regions.
Therefore, the polarization parameters obtained in Collins-Soper frame are
expected to be similar to those obtained in the helicity frame,
except at high \pt and low $y$ bins.
Calculating the frame-invariant variable, according to Eq.~(\ref{invpar}), the
measurements performed in the two frames are in agreement within the
uncertainty.
\par
\begin{figure}[!htbp]
  \begin{minipage}[t]{0.5\textwidth}
    \centering
    \includegraphics[width=1.0\textwidth]{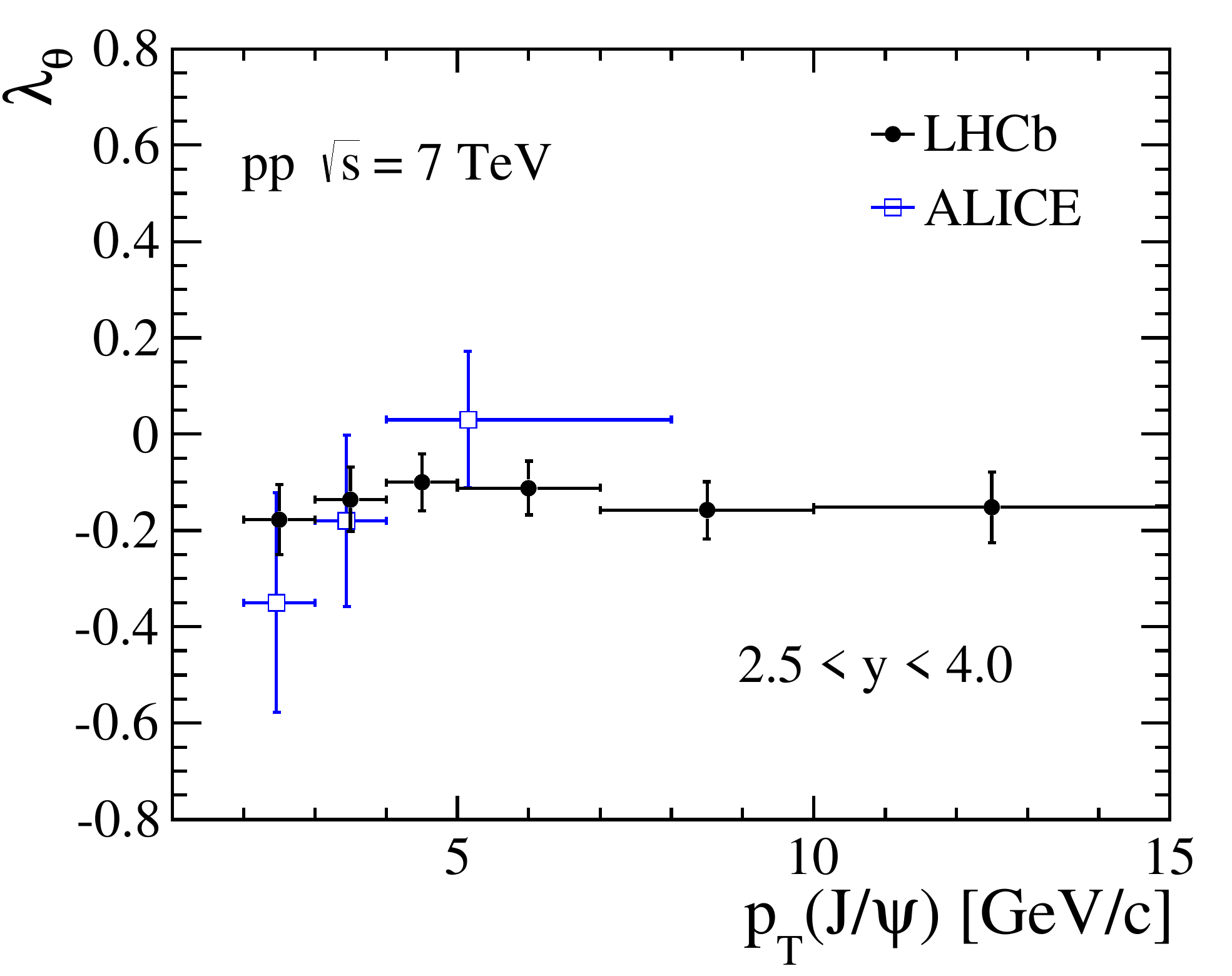}
  \end{minipage}
  \begin{minipage}[t]{0.5\textwidth}
    \centering
    \includegraphics[width=1.0\textwidth]{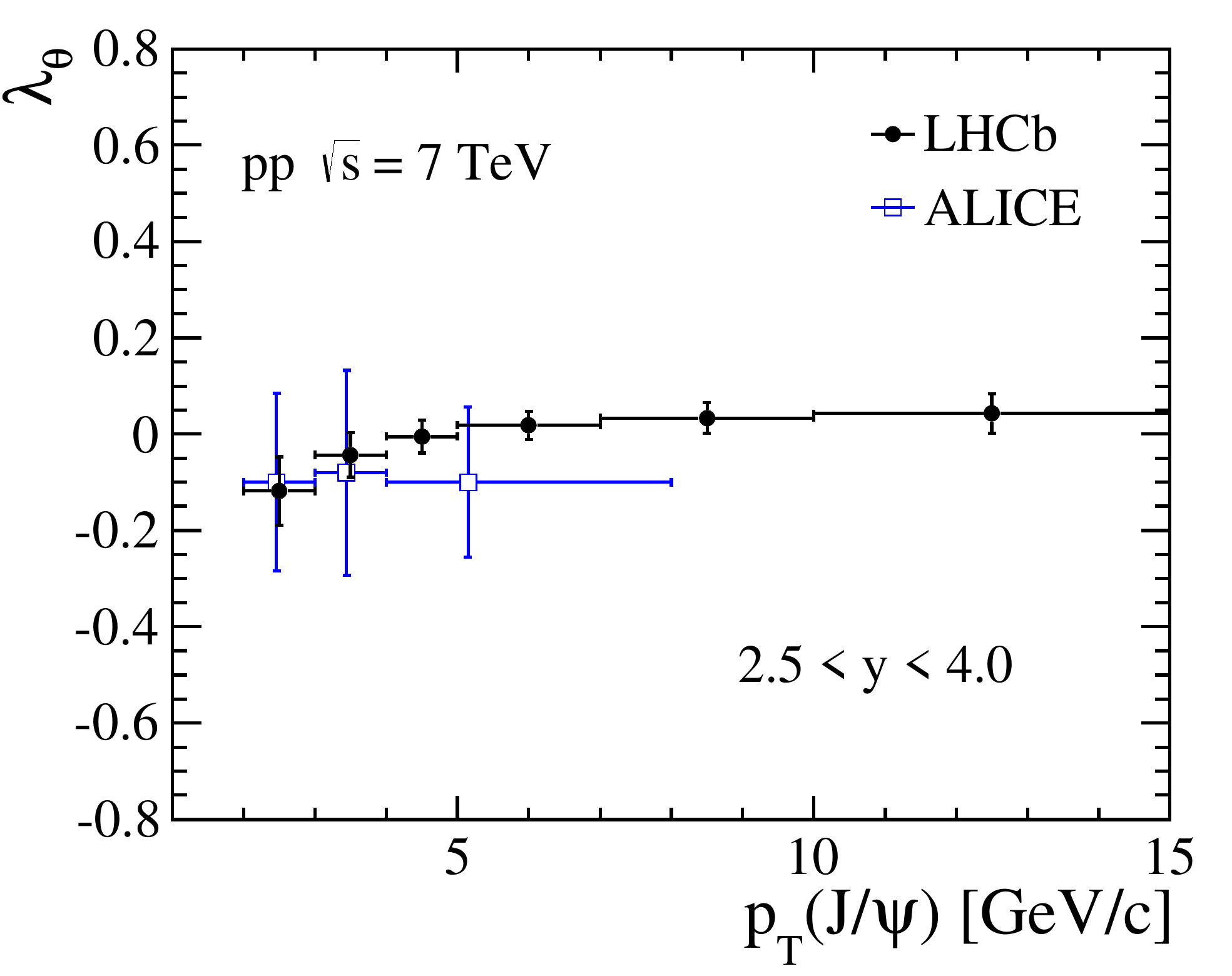}
  \end{minipage}
  \caption{\small
    Comparison of LHCb and ALICE results for $\lambda_{\theta}$ in different
    \pt bins integrating over the rapidity range $2.5 < y < 4.0$ in
    (\textit{left}) the helicity frame and (\textit{right}) the Collins-Soper
    frame.
    Error bars represent the statistical and systematic uncertainties added in
    quadrature.}\label{lthetaAlice}
\end{figure}

The results can be compared to those obtained by other experiments at
different valuses of $\sqrt{s}$.
Measurements by CDF \cite{CDFresults_jpsi_pol}, PHENIX \cite{RHICresults} and
HERA-B \cite{HERAresults}, also favour a negative value for $\lambda _\theta$.
The HERA-B experiment has also published results on $\lambda _{\phi}$ and
$\lambda _{\theta \phi}$, which are consistent with zero.
At the LHC, the ALICE~\cite{ALICEresults} and the CMS~\cite{Chatrchyan:2013cla} collaboration studied the \jpsi polarization in $pp$
collisions at $\sqrt{s} = 7$~\tev.
The CMS results, determined in a different kinematic range, disfavour large transverse or longitudinal polarizations.
The analysis by ALICE is based on the $\cos\theta$ and $\phi$ projections and thus only
determines $\lambda_\theta$ and $\lambda_\phi$.
Furthermore it also includes \jpsi mesons from \bquark -hadron decays.
The measurement has been performed in bins of \jpsi transverse momentum
integrating over the rapidity in a range very similar to that of
LHCb, being $2 < \pt < 8$~\gevc and \mbox{$2.5 < y < 4.0$}.
To compare our results with the ALICE measurements, averages over the $y$
region are used for the different \pt bins and good agreement is found for
$\lambda_{\theta}$ and $\lambda _{\phi}$.
The comparison for $\lambda _{\theta}$ is shown in Fig.~\ref{lthetaAlice} for
the helicity and Collins-Soper frames, respectively.

\begin{figure}[!htbp]
  \begin{center}
   \includegraphics[angle=0,width=0.8\textwidth]{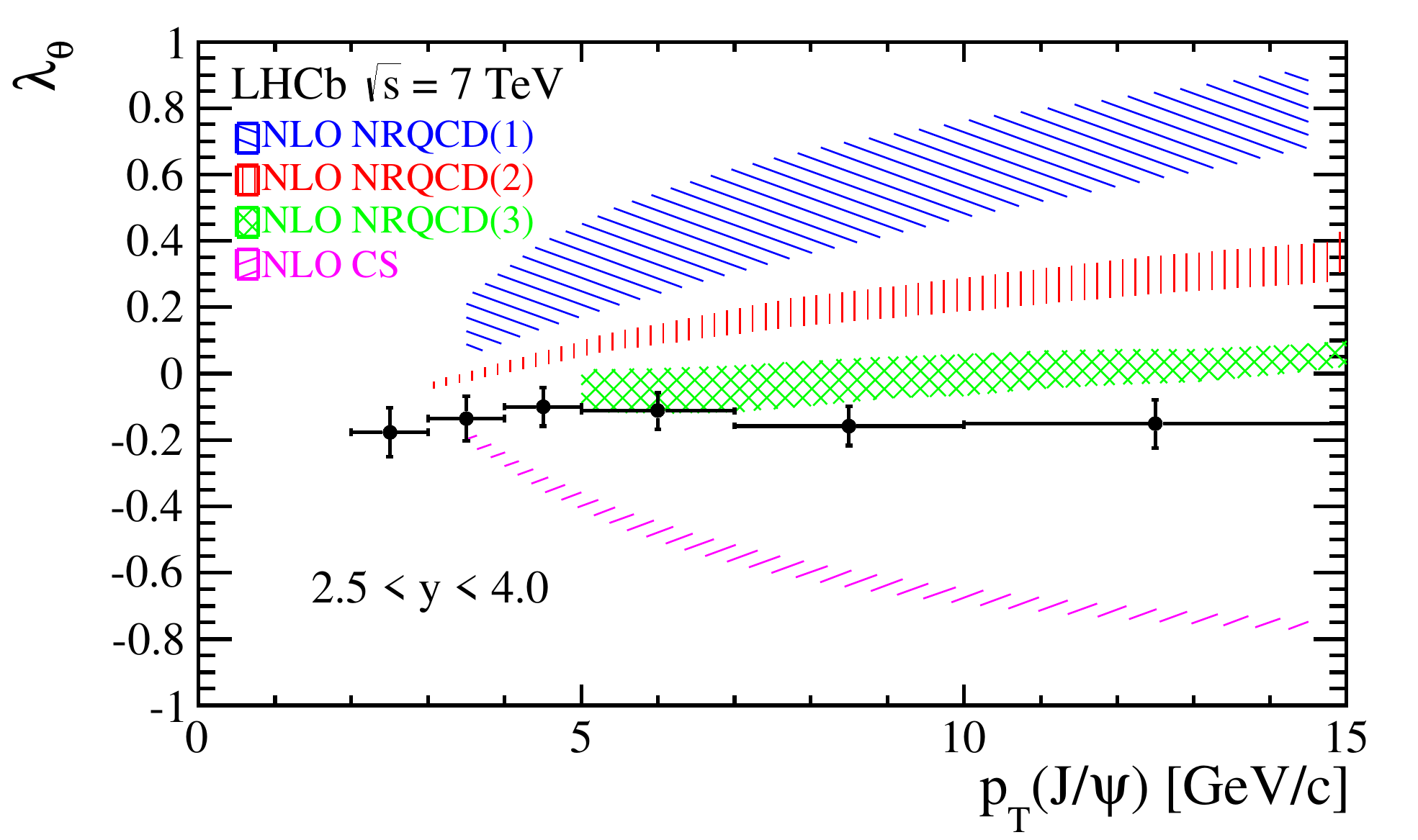}
   \caption{\small
     Comparison of LHCb prompt \jpsi polarization measurements of
     $\lambda_{\theta}$ with direct NLO color singlet
     (magenta diagonal lines \cite{TheoryMathias}) and three different NLO NRQCD 
     (blue diagonal lines (1) \cite{TheoryMathias}, red vertical lines (2)
     \cite{Gong:2012ug} and green hatched (3) \cite{Chao:2012iv,*Shao2012fs}) predictions as a
     function of the \pt of the \jpsi meson in the rapidity range $2.5<y<4.0$
     in the helicity frame.}
   \label{CompareWithTheory}
  \end{center}
\end{figure}

In Fig.~\ref{CompareWithTheory} our measurements of $\lambda_\theta$ are
compared with the NLO CSM~\cite{TheoryMathias} and NRQCD predictions of
Refs.~\cite{TheoryMathias}, \cite{Gong:2012ug} and \cite{Chao:2012iv,*Shao2012fs}.
The comparison is done in the helicity frame and as a function of the \pt of the
\jpsi meson (integrating over $ 2.5 < y < 4.0$).
The theoretical calculations in Refs.~\cite{TheoryMathias}, \cite{Gong:2012ug}
and \cite{Chao:2012iv,*Shao2012fs} use different selections of experimental data to evaluate the non-perturbative
matrix elements.
Our results are not in agreement with the CSM predictions and the best
agreement is found between the measured values and the NRQCD predictions of
Ref.~\cite{Chao:2012iv,*Shao2012fs}.
It should be noted that our analysis includes a contribution from feed-down,
while the theoretical computations from CSM and NRQCD~\cite{TheoryMathias} do
not include feed-down from excited states.
It is known that, among all the feed-down contributions to prompt \jpsi production from
higher charmonium states, the contribution from \chic mesons can be quite important (up to 30\%) and that 
\psitwos mesons also can give a sizable contribution~\cite{Faccioli:2012nv,Gong:2012ug,Chao:2012iv,*Shao2012fs}, 
depending on the yields and their polarizations. 
The NLO NRQCD calculations~\cite{Gong:2012ug,Chao:2012iv,*Shao2012fs} include
the feed-down from \chic and \psitwos mesons.

\section[Cross-section update]{Update of the $\mathbf{\jpsi}$ cross-section measurement}
\label{crosssecref}
The \jpsi cross-section in $pp$ collisions at $\sqrt{s} = 7~\tev$ was previously measured by LHCb in 14 bins of
\pt and five bins of $y$ of the \jpsi meson~\cite{LHCb-PAPER-2011-003}.
The uncertainty on the prompt cross-section measurement is dominated by the
unknown \jpsi polarization, resulting in uncertainties of up to 20\%:
\begin{equation*}
\sigma _{\mathrm{prompt}} (2 < y < 4.5, \pt < 14 \gevc) =  10.52 \pm 0.04 \pm 1.40 \, ^{+1.64}_{-2.20}~\mub
\end{equation*}
where the first uncertainty is statistical, the second is systematic and the third one is due to the unknown polarization.
\par
The previous measurement of the prompt \jpsi cross-section can be updated in the range of the polarization analysis, $2<\pt < 14$~\gevc and $2.0 < y < 4.5$,
by applying the measured polarization and its uncertainty to the efficiency calculation in the cross-section measurement.
To re-evaluate the \jpsi production cross-section, the same data sample, trigger and selection requirements as in Ref.~\cite{LHCb-PAPER-2011-003} are used. 
Technically the polarization correction is done by reweighting the muon angular distribution of a simulated sample of unpolarized \decay{\jpsi}{\mup \mun} events to reproduce the expected distribution, according to Eq.~(\ref{theory1}), for polarized \jpsi mesons. The
polarization parameters $\lambda_\theta$, $\lambda_{\theta\phi}$ and $\lambda_\phi$ are set to the measured values, quoted in Table \ref{tab:PolarizationResultHX} for each  bin of \pt and $y$ of the \jpsi meson.
\par
\begin{figure}
  \centering
  \includegraphics[width=.8\textwidth]{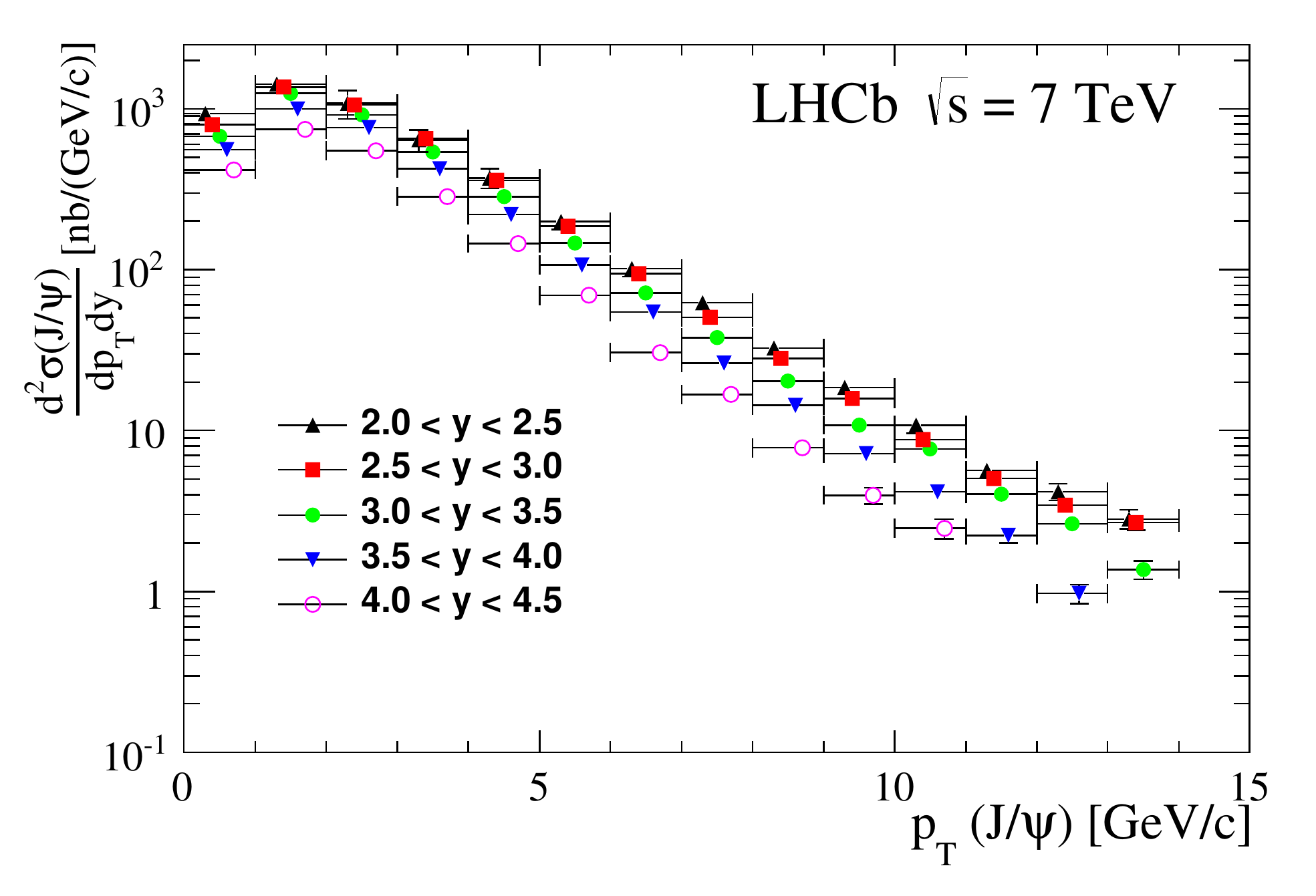}
  \label{XSec}
  \caption{\small
    Differential cross-section of prompt \jpsi production as a function of
    \pt and in bins of $y$.
    The vertical error bars show the quadratic sum of the statistical and
    systematic uncertainties.}
\end{figure}

In addition to the polarization update, the uncertainties on the
luminosity determination and on the track reconstruction efficiency are
updated to take into account the improvements described in
Refs.~\cite{LHCb-PAPER-2011-015, tracking}.
For the tracking efficiency it is possible to reduce the systematic
uncertainty to 3\%, compared to an 8\% uncertainty
assigned in the original measurement\cite{LHCb-PAPER-2011-003}. Taking advantage of the improvements described in~\cite{LHCb-PAPER-2011-015} the uncertainty due to the luminosity measurement has been reduced from the 10\%, quoted in~\cite{LHCb-PAPER-2011-003} to the 3.5\%.
The results obtained for the double-differential cross-section are shown in
Fig.~\ref{XSec} and reported in Table~\ref{tab:XSecAll}.
The integrated cross-section in the kinematic range of the polarization
analysis, 2 $< \pt < 14$~\gevc and $2.0 < y < 4.5$, is
\begin{equation*}
  \sigma _{\mathrm{prompt}} (2 < y < 4.5, 2 < \pt < 14 \gevc) = 4.88 \pm 0.01 \pm 0.27 \pm 0.12 \;  \mub
  \label{polrange}
\end{equation*}
and for the range $\pt < 14$ \gevc and $2.0 < y < 4.5$, it is
\begin{equation*}
  \sigma _{\mathrm{prompt}} (2 < y < 4.5, \pt < 14 \gevc) = 9.46 \pm 0.04 \pm 0.53 \, ^{+0.86}_{-1.10} \; \mub.
  \label{oldrange}
\end{equation*}
For the two given cross-section measurements, the first uncertainty is statistical, the second is systematic, while the third arises from the remaining uncertainty
due to the polarization measurement and is evaluated using simulated event
samples.
For the \pt range $\pt < 2$~\gevc, where no polarization measurement exists, we
assume zero polarization and assign as systematic uncertainty the
difference between  the zero polarization  hypothesis and fully
transverse (upper values) or  fully longitudinal
(lower values) polarization.
For $\pt > 2\gevc$ the uncertainties on the polarization measurement coming from the various sources are propagated to the cross-section measurement fluctuating the values of the polarization parameters in Eq.~\ref{theory1} with a Gaussian width equal to one standard deviation. The relative uncertainty due to the polarization effect on the integrated cross-section in $2 < \pt < 14$~\gevc and $2.0 < y < 4.5$ is $2.4 \%$. The relative uncertainty on the integrated cross-section in the range of Ref.~\cite{LHCb-PAPER-2011-003}, $\pt < 14$~\gevc and $2.0 < y < 4.5$, is reduced to $12 \%$ (lower polarization uncertainty) and to $9\%$ (upper polarization uncertainty) with respect to the value published in Ref.~\cite{LHCb-PAPER-2011-003}.

\section{Conclusion} \label{conclusion}
A measurement of the prompt \jpsi polarization obtained with $pp$ collisions at $\sqrt{s}= 7$~\tev, performed using a dataset corresponding to an integrated luminosity of 0.37~\invfb, is presented.
The data have been collected by the LHCb experiment in the early 2011.
The polarization parameters ($\lambda_{\theta}, \lambda_{\theta\phi},
\lambda_{\phi}$) are determined by studying the angular distribution of the
two muons from the decay \decay{\jpsi}{\mup \mun} with respect to the polar
and azimuthal angle defined in the helicity frame.
The measurement is performed in five bins of \jpsi rapidity $y$ and six
bins of \jpsi transverse momentum \pt in the kinematic range $2 < p_{\mathrm{T}} < 15 \gevc$ and $2.0 < y < 4.5$.
\par
The results for $\lambda _{\theta}$ indicate a small longitudinal polarization while the results for $\lambda _{\theta \phi}$ and $\lambda_{\phi}$ are consistent
with zero.
Although a direct comparison is not possible due to the different collision energies and analysis ranges, the measurements performed by CDF \cite{CDFresults_jpsi_pol}, PHENIX \cite{RHICresults}, 
HERA-B \cite{HERAresults} and CMS~\cite{Chatrchyan:2013cla} show no significant transverse or longitudinal
polarization.
Good agreement has also been found with ALICE measurements
\cite{ALICEresults}, performed in a \pt and rapidity range very similar to
that explored by LHCb.
\par
Our results, that are obtained for prompt \jpsi production, including the
feed-down from higher excited states, contradict the CSM predictions for
direct \jpsi production, both in the size of the polarization parameters and
the $\pt$ dependence.
Concerning the NRQCD models, predictions from Ref.~\cite{Chao:2012iv,*Shao2012fs} give the
best agreement with the LHCb measurement.
\par
This evaluation of the polarization is used to update the measurement of
the integrated \jpsi production cross-section \cite{LHCb-PAPER-2011-003} in the range $\pt < 14 \gevc$ and $2.0 < y < 4.5$, resulting in a reduction of the corresponding systematic uncertainty to $9\%$ and $12\%$. The result is
\begin{equation*}
  \sigma _{\mathrm{prompt}} (2 < y < 4.5, \pt < 14 \gevc) = 9.46 \pm 0.04 \pm 0.53 \, ^{+0.86}_{-1.10} \; \mub.
  \label{oldrange}
\end{equation*}
The integrated cross-section has also been measured in the polarization analysis range $2 < \pt < 14 \gevc$ and $2.0 < y < 4.5$:
\begin{equation*}
  \sigma _{\mathrm{prompt}} (2 < y < 4.5, 2 < \pt < 14 \gevc) = 4.88 \pm 0.01 \pm 0.27 \pm 0.12 \;  \mub.
  \label{polrange}
\end{equation*}
with an uncertainty due to polarization of $2.4\%$.

\section*{Acknowledgements}
\noindent 
We wish to thank M. Butenschoen, B. Gong and  Y.-Q. Ma for providing 
us the theoretical calculations and helpful discussions. 
We are grateful for fruitful discussions with S. P. Baranov.
We express our gratitude to our colleagues in the CERN
accelerator departments for the excellent performance of the LHC. We
thank the technical and administrative staff at the LHCb
institutes. We acknowledge support from CERN and from the national
agencies: CAPES, CNPq, FAPERJ and FINEP (Brazil); NSFC (China);
CNRS/IN2P3 and Region Auvergne (France); BMBF, DFG, HGF and MPG
(Germany); SFI (Ireland); INFN (Italy); FOM and NWO (The Netherlands);
SCSR (Poland); MEN/IFA (Romania); MinES, Rosatom, RFBR and NRC
``Kurchatov Institute'' (Russia); MinECo, XuntaGal and GENCAT (Spain);
SNSF and SER (Switzerland); NAS Ukraine (Ukraine); STFC (United
Kingdom); NSF (USA). We also acknowledge the support received from the
ERC under FP7. The Tier1 computing centres are supported by IN2P3
(France), KIT and BMBF (Germany), INFN (Italy), NWO and SURF (The
Netherlands), PIC (Spain), GridPP (United Kingdom). We are thankful
for the computing resources put at our disposal by Yandex LLC
(Russia), as well as to the communities behind the multiple open
source software packages that we depend on.

\clearpage

{\noindent\bf\Large Appendices}

\appendix

\begin{landscape}
  \begin{table}
    \caption{\small
      Measured \jpsi polarization parameters in bins of \pt and $y$
      in the helicity frame.
      The first uncertainty is statistical (from the fit and the background
      subtraction) while the second is the systematic uncertainty.}
    \label{tab:PolarizationResultHX}
    \begin{center}
      \small
      \begin{tabular}{ccc@{$\pm$}c@{$\pm$}cc@{$\pm$}c@{$\pm$}cc@{$\pm$}c@{$\pm$}cc@{$\pm$}c@{$\pm$}cc@{$\pm$}c@{$\pm$}c}
        \toprule[1.0pt]
        \pt (\gevc) & $\lambda$ & \multicolumn{3}{c}{$2.0<y<2.5$} & \multicolumn{3}{c}{$2.5<y<3.0$} & \multicolumn{3}{c}{$3.0<y<3.5$} & \multicolumn{3}{c}{$3.5<y<4.0$} & \multicolumn{3}{c}{$4.0<y<4.5$}  \\
        \midrule[1.0pt]
        \midrule[1.0pt]
&$\lambda_{\theta}$&-0.306&0.095&0.288
&-0.207&0.010&0.101
&-0.169&0.006&0.066
&-0.161&0.005&0.059
&-0.081&0.008&0.092
\\2-3&$\lambda_{\theta\phi}$&0.057&0.052&0.114
&-0.055&0.004&0.039
&-0.054&0.003&0.034
&0.004&0.003&0.043
&0.052&0.006&0.050
\\
&$\lambda_{\phi}$&0.034&0.016&0.075
&0.023&0.003&0.043
&0.009&0.002&0.027
&0.036&0.003&0.026
&0.048&0.005&0.041
\\
        \midrule[1.0pt]
&$\lambda_{\theta}$&-0.419&0.073&0.218
&-0.077&0.010&0.100
&-0.173&0.006&0.056
&-0.149&0.006&0.054
&-0.125&0.010&0.086
\\3-4&$\lambda_{\theta\phi}$&-0.055&0.044&0.094
&-0.024&0.004&0.030
&-0.029&0.003&0.023
&0.022&0.003&0.026
&0.045&0.005&0.046
\\
&$\lambda_{\phi}$&0.021&0.016&0.045
&-0.014&0.003&0.018
&-0.002&0.003&0.019
&0.029&0.003&0.025
&0.013&0.006&0.034
\\
        \midrule[1.0pt]
&$\lambda_{\theta}$&-0.390&0.056&0.174
&-0.022&0.010&0.077
&-0.149&0.007&0.050
&-0.129&0.007&0.055
&-0.158&0.012&0.099
\\4-5&$\lambda_{\theta\phi}$&-0.059&0.037&0.075
&-0.013&0.004&0.029
&-0.037&0.004&0.023
&0.003&0.004&0.026
&0.078&0.007&0.048
\\
&$\lambda_{\phi}$&0.032&0.015&0.038
&-0.004&0.003&0.015
&-0.009&0.003&0.017
&0.025&0.004&0.022
&-0.015&0.008&0.031
\\

        \midrule[1.0pt]
&$\lambda_{\theta}$&-0.126&0.037&0.133
&-0.072&0.009&0.067
&-0.158&0.007&0.048
&-0.104&0.008&0.055
&-0.045&0.013&0.098
\\5-7&$\lambda_{\theta\phi}$&-0.051&0.024&0.064
&-0.010&0.004&0.026
&0.007&0.004&0.022
&-0.022&0.005&0.026
&0.005&0.008&0.053
\\
&$\lambda_{\phi}$&-0.016&0.010&0.031
&-0.014&0.003&0.012
&-0.035&0.003&0.014
&0.027&0.003&0.018
&0.030&0.007&0.026
\\

        \midrule[1.0pt]
&$\lambda_{\theta}$&0.009&0.037&0.120
&-0.217&0.012&0.064
&-0.162&0.011&0.055
&-0.042&0.013&0.066
&-0.057&0.020&0.100
\\7-10&$\lambda_{\theta\phi}$&0.027&0.023&0.048
&-0.016&0.005&0.026
&0.029&0.005&0.022
&0.006&0.007&0.028
&-0.005&0.012&0.053
\\
&$\lambda_{\phi}$&0.003&0.010&0.024
&-0.008&0.004&0.011
&-0.025&0.004&0.013
&0.007&0.005&0.016
&0.034&0.010&0.027
\\

        \midrule[1.0pt]
&$\lambda_{\theta}$&-0.248&0.047&0.115
&-0.267&0.020&0.075
&-0.040&0.022&0.077
&-0.076&0.028&0.082
&-0.089&0.046&0.115
\\10-15&$\lambda_{\theta\phi}$&-0.088&0.027&0.054
&-0.012&0.009&0.028
&0.018&0.010&0.023
&0.010&0.014&0.035
&-0.043&0.025&0.042
\\
&$\lambda_{\phi}$&0.009&0.014&0.029
&0.008&0.007&0.013
&-0.018&0.009&0.017
&-0.014&0.012&0.019
&-0.027&0.021&0.040
\\
        \bottomrule[1.0pt]

      \end{tabular}
    \end{center}
  \end{table}
\end{landscape}

\begin{landscape}
  \begin{table}
    \caption{\small
      Measured \jpsi polarization parameters in bins of \pt and
      $y$ in Collins-Soper frame.
      The first uncertainty is statistical (from the fit and the background
      subtraction) while the second is the systematic uncertainty.}
    \label{tab:PolarizationResultCS}
    \begin{center}
      \small
      \begin{tabular}{ccc@{$\pm$}c@{$\pm$}cc@{$\pm$}c@{$\pm$}cc@{$\pm$}c@{$\pm$}cc@{$\pm$}c@{$\pm$}cc@{$\pm$}c@{$\pm$}c}
        \toprule[1.0pt]
        \pt (\gevc) & $\lambda$ & \multicolumn{3}{c}{$2.0<y<2.5$} & \multicolumn{3}{c}{$2.5<y<3.0$} & \multicolumn{3}{c}{$3.0<y<3.5$} & \multicolumn{3}{c}{$3.5<y<4.0$} & \multicolumn{3}{c}{$4.0<y<4.5$}  \\
        \midrule[1.pt]
        \midrule[1.pt]
        
  &$\lambda_{\theta}$&-0.305&0.118&0.338
&-0.176&0.009&0.108
&-0.130&0.004&0.058
&-0.051&0.005&0.067
&-0.043&0.011&0.085
\\2-3&$\lambda_{\theta\phi}$&0.152&0.044&0.158
&0.114&0.006&0.058
&0.102&0.004&0.035
&0.098&0.003&0.036
&0.037&0.005&0.050
\\
&$\lambda_{\phi}$&-0.031&0.011&0.125
&0.014&0.003&0.059
&0.008&0.002&0.038
&-0.001&0.002&0.031
&-0.005&0.003&0.036
\\

        \midrule[1.pt]
&$\lambda_{\theta}$&-0.180&0.086&0.215
&-0.076&0.007&0.067
&-0.064&0.004&0.034
&0.017&0.005&0.042
&-0.001&0.011&0.070
\\3-4&$\lambda_{\theta\phi}$&0.223&0.042&0.095
&0.090&0.006&0.047
&0.109&0.004&0.031
&0.081&0.004&0.032
&0.015&0.006&0.049
\\
&$\lambda_{\phi}$&-0.070&0.014&0.065
&-0.027&0.004&0.036
&-0.033&0.003&0.028
&-0.017&0.004&0.026
&-0.049&0.005&0.040
\\

        \midrule[1.pt]
&$\lambda_{\theta}$&-0.084&0.068&0.171
&-0.000&0.007&0.040
&-0.035&0.005&0.030
&0.031&0.006&0.037
&0.051&0.012&0.071
\\4-5&$\lambda_{\theta\phi}$&0.240&0.041&0.092
&0.067&0.006&0.041
&0.081&0.004&0.027
&0.065&0.004&0.030
&-0.028&0.008&0.052
\\
&$\lambda_{\phi}$&-0.104&0.017&0.055
&-0.042&0.005&0.032
&-0.050&0.005&0.027
&-0.033&0.005&0.029
&-0.095&0.007&0.047
\\

        \midrule[1.pt]
&$\lambda_{\theta}$&-0.110&0.037&0.081
&0.008&0.006&0.032
&0.005&0.005&0.027
&0.054&0.006&0.033
&0.089&0.012&0.072
\\5-7&$\lambda_{\theta\phi}$&0.160&0.029&0.070
&0.056&0.005&0.032
&0.041&0.004&0.023
&0.063&0.004&0.028
&-0.000&0.008&0.053
\\
&$\lambda_{\phi}$&-0.068&0.014&0.051
&-0.056&0.005&0.031
&-0.085&0.005&0.026
&-0.051&0.005&0.031
&-0.056&0.008&0.052
\\

        \midrule[1.pt]
&$\lambda_{\theta}$&0.079&0.032&0.061
&0.035&0.009&0.035
&0.032&0.009&0.030
&0.031&0.011&0.036
&0.072&0.020&0.071
\\7-10&$\lambda_{\theta\phi}$&0.014&0.028&0.061
&0.073&0.006&0.026
&0.036&0.005&0.023
&0.022&0.007&0.029
&0.007&0.013&0.045
\\
&$\lambda_{\phi}$&-0.074&0.018&0.053
&-0.078&0.007&0.032
&-0.076&0.007&0.029
&-0.027&0.009&0.036
&-0.022&0.014&0.055
\\

        \midrule[1.pt]
&$\lambda_{\theta}$&0.064&0.037&0.076
&0.099&0.016&0.046
&-0.004&0.018&0.044
&-0.009&0.024&0.050
&0.019&0.042&0.086
\\10-15&$\lambda_{\theta\phi}$&0.105&0.033&0.057
&0.070&0.010&0.024
&0.004&0.010&0.024
&0.021&0.014&0.028
&0.033&0.026&0.041
\\
&$\lambda_{\phi}$&-0.093&0.026&0.059
&-0.108&0.013&0.040
&-0.024&0.013&0.040
&-0.024&0.017&0.048
&-0.084&0.030&0.064
\\
        \bottomrule[1.pt]

      \end{tabular}
    \end{center}
  \end{table}
\end{landscape}

\begin{landscape}
\begin{table}
  \caption{\small
    Double-differential cross-section $d ^{2} \sigma/ d\pt \, dy$ in nb/(\gevc) for prompt \jpsi production in bins of \pt and $y$, with statistical, systematic and polarization uncertainties.}\label{tab:XSecAll}
  \begin{center}
  \small
      \begin{tabular}{llllllllll} 
        \toprule[1.0pt]
\pt (\gevc) & $2.0 < y < 2.5$ & & $2.5 < y < 3.0$ & & $3.0 < y < 3.5$ & \\     
\midrule[1.0pt]
\midrule[1.0pt]   
2-3& 1083 $\pm$ 18 $\pm$ 64 $\pm$ 210&  & 1055 $\pm$ 8 $\pm$ 61 $\pm$ 47&  & 918 $\pm$ 6 $\pm$ 53 $\pm$ 28\\
3-4& 639 $\pm$ 9 $\pm$ 41 $\pm$ 93&  & 653 $\pm$ 5 $\pm$ 39 $\pm$ 28&  & 541 $\pm$ 4 $\pm$ 32 $\pm$ 17\\
4-5 & 370 $\pm$ 5 $\pm$ 24 $\pm$ 46&  & 359.1 $\pm$ 3.1 $\pm$ 22.3 $\pm$ 14.1&  & 285.1 $\pm$ 2.4 $\pm$ 17.7 $\pm$ 8.5\\
5-6 & 199.0 $\pm$ 3.0 $\pm$ 13.8 $\pm$ 17.4&  & 185.9 $\pm$ 2.0 $\pm$ 12.2 $\pm$ 6.2&  & 146.4 $\pm$ 1.7 $\pm$ 9.3 $\pm$ 4.2\\
6-7 & 101.2 $\pm$ 1.9 $\pm$ 7.3 $\pm$ 8.0&  & 94.1 $\pm$ 1.3 $\pm$ 6.4 $\pm$ 2.9&  & 71.7 $\pm$ 1.1 $\pm$ 4.8 $\pm$ 1.9\\
7-8 & 62.2 $\pm$ 1.4 $\pm$ 4.1 $\pm$ 4.6&  & 50.6 $\pm$ 0.9 $\pm$ 3.7 $\pm$ 1.7&  & 37.8 $\pm$ 0.7 $\pm$ 2.4 $\pm$ 1.2\\
8-9 & 32.5 $\pm$ 0.9 $\pm$ 2.1 $\pm$ 2.2&  & 28.1 $\pm$ 0.7 $\pm$ 1.8 $\pm$ 0.9&  & 20.3 $\pm$ 0.5 $\pm$ 1.3 $\pm$ 0.6\\
9-10 & 18.5 $\pm$ 0.7 $\pm$ 1.2 $\pm$ 1.3&  & 15.8 $\pm$ 0.5 $\pm$ 1.0 $\pm$ 0.5&  & 10.8 $\pm$ 0.4 $\pm$ 0.7 $\pm$ 0.3\\
10-11 & 10.8 $\pm$ 0.5 $\pm$ 0.7 $\pm$ 0.9&  & 8.7 $\pm$ 0.4 $\pm$ 0.6 $\pm$ 0.3&  & 7.70 $\pm$ 0.34 $\pm$ 0.50 $\pm$ 0.31\\
11-12 & 5.65 $\pm$ 0.32 $\pm$ 0.37 $\pm$ 0.41&  & 5.04 $\pm$ 0.26 $\pm$ 0.32 $\pm$ 0.18&  & 4.03 $\pm$ 0.23 $\pm$ 0.26 $\pm$ 0.13\\
12-13 & 4.16 $\pm$ 0.27 $\pm$ 0.27 $\pm$ 0.32&  & 3.42 $\pm$ 0.23 $\pm$ 0.22 $\pm$ 0.14&  & 2.64 $\pm$ 0.18 $\pm$ 0.17 $\pm$ 0.09\\
13-14 & 2.82 $\pm$ 0.26 $\pm$ 0.19 $\pm$ 0.21&  & 2.68 $\pm$ 0.20 $\pm$ 0.17 $\pm$ 0.11&  & 1.37 $\pm$ 0.15 $\pm$ 0.09 $\pm$ 0.06\\
\midrule[1.0pt]
\pt (\gevc) & $3.5 < y < 4.0$ & & $4.0 < y < 4.5$ & \\    
\midrule[1.0pt]
2-3 & 762 $\pm$ 5 $\pm$ 46 $\pm$ 23&  & 549 $\pm$ 5 $\pm$ 36 $\pm$ 27\\
3-4 & 422.9 $\pm$ 3.4 $\pm$ 26.2 $\pm$ 12.9&  & 284 $\pm$ 3 $\pm$ 19 $\pm$ 16\\
4-5 & 219.1 $\pm$ 2.3 $\pm$ 13.9 $\pm$ 6.7&  & 145.4 $\pm$ 2.4 $\pm$ 9.2 $\pm$ 8.7\\
5-6 & 107.2 $\pm$ 1.4 $\pm$ 7.5 $\pm$ 3.2&  & 69.2 $\pm$ 1.5 $\pm$ 4.4 $\pm$ 3.5\\
6-7 & 54.6 $\pm$ 1.0 $\pm$ 3.5 $\pm$ 1.6&  & 30.6 $\pm$ 1.0 $\pm$ 1.9 $\pm$ 1.4\\
7-8 & 26.2 $\pm$ 0.6 $\pm$ 1.7 $\pm$ 0.9&  & 16.71 $\pm$ 0.69 $\pm$ 1.06 $\pm$ 0.92\\
8-9 & 14.3 $\pm$ 0.5 $\pm$ 0.9 $\pm$ 0.5&  & 7.78 $\pm$ 0.43 $\pm$ 0.49 $\pm$ 0.39\\
9-10 & 7.18 $\pm$ 0.32 $\pm$ 0.46 $\pm$ 0.22&  & 3.96 $\pm$ 0.31 $\pm$ 0.25 $\pm$ 0.24\\
10-11 & 4.15 $\pm$ 0.24 $\pm$ 0.27 $\pm$ 0.18&  & 2.47 $\pm$ 0.25 $\pm$ 0.16 $\pm$ 0.18\\
11-12 & 2.24 $\pm$ 0.17 $\pm$ 0.14 $\pm$ 0.08&  & -\\
12-13 & 0.97 $\pm$ 0.11 $\pm$ 0.06 $\pm$ 0.04&  & -\\
13-14 & - &  & -\\     

        \bottomrule[1.pt]
      \end{tabular}
  \end{center}
\end{table}
\end{landscape}

\addcontentsline{toc}{section}{References}
\bibliographystyle{LHCb}
\bibliography{main}

\end{document}